\documentclass[pra,letterpaper,twocolumn,showpacs,superscriptaddress]{revtex4}
\usepackage{graphicx,psfrag,amsmath,amssymb,amsfonts,bbm,latexsym,color,dcolumn}

\begin{document}

\title{Cooling and thermometry of atomic Fermi gases}

\author{Roberto Onofrio}

\affiliation{{\mbox Dipartimento di Fisica e Astronomia ``Galileo Galilei'', Universit\`a di Padova, \\
Via Marzolo 8, Padova 35131, Italy}}

\affiliation{{\mbox Department of Physics and Astronomy, Dartmouth College, \\ 6127 Wilder Laboratory, 
Hanover, NH 03755, USA}}

\begin{abstract}
We review the status of cooling techniques aimed at achieving the deepest quantum degeneracy for atomic Fermi gases. 
We first discuss some physical motivations, providing a quantitative assessment of the need for deep quantum degeneracy 
in relevant physics cases, such as the search for unconventional superfluid states. Attention is then focused on the most 
widespread technique to reach deep quantum degeneracy for Fermi systems, sympathetic cooling of Bose-Fermi mixtures, 
organizing the discussion according to the specific species involved. Various proposals to circumvent some of the 
limitations on achieving the deepest Fermi degeneracy, and their experimental realizations, are then reviewed. 
Finally, we discuss the extension of these techniques to optical lattices and the implementation of precision thermometry 
crucial to the understanding of the phase diagram of classical and quantum phase transitions in Fermi gases.
\end{abstract}

\pacs{03.75.Ss, 05.30.Fk, 07.20.Dt, 37.10.De, 67.60.Bc}

\maketitle

\noindent
{\bf Keywords:} ultracold Fermi gases, Fermi-Bose mixtures, superfluidity phenomena, atomic trapping, thermometry

\vspace{0.4cm}

\noindent
{\bf Contents}
\vspace{0.4cm}

I. Introduction \hfill{1}

II. Searching for exotic superfluid states \hfill{3}

III. Generalities of fermion cooling \hfill{3}

IV. Dual evaporative cooling of fermions \hfill{4}

V. Fermi-Bose mixtures \hfill{6}

\hspace{1.0cm} A. ${}^6$Li-${}^7$Li \hfill{6}

\hspace{1.0cm} B. ${}^6$Li-${}^{23}$Na \hfill{8}

\hspace{1.0cm} C. ${}^{40}$K-${}^{87}$Rb \hfill{8}

\hspace{1.0cm} D. ${}^6$Li-${}^{87}$Rb \hfill{9}

\hspace{1.0cm} E. ${}^{3}$He-${}^{4}$He \hfill{9}

\hspace{1.0cm} F. ${}^{171}$Yb-${}^{174}$Yb and ${}^{173}$Yb-${}^{174}$Yb \hfill{9}

\hspace{1.0cm} G. ${}^{87}$Sr-${}^{84}$Sr \hfill{10}

\hspace{1.0cm} H. ${}^{6}$Li-${}^{174}$Yb \hfill{10}

\hspace{1.0cm} I. ${}^{171}$Yb-${}^{87}$Rb \hfill{10}

\hspace{1.0cm} J. ${}^{40}$K-${}^{23}$Na \hfill{10}

\hspace{1.0cm} K.  ${}^{6}$Li-${}^{41}$K \hfill{11}

\hspace{1.0cm} L. Current experimental situation  \hfill{11}

VI. Experimental techniques to achieve deeper

\hspace{1.0cm}Fermi degeneracy \hfill{11}

\hspace{1.0cm} A. Species-selective trapping \hfill{12}

\hspace{1.0cm} B. Frictionless adiabatic expansion 

\hspace{1.5cm}or compression \hfill{15}

\hspace{1.0cm} C. Heat capacity matching by reduced 

\hspace{1.5cm}dimensionality \hfill{19}

\hspace{1.0cm} D. All-optical cooling techniques \hfill{20}

VII. Optical lattices \hfill{22}

VIII. General issues in precision thermometry \hfill{24}

IX. Conclusions \hfill{25}

References \hfill{27}

\vspace{1.0cm}

\section{\bf{Introduction}}
The study of superfluid phenomena has been, since its very inception, crucial in providing experimental 
insights into the quantum description of macroscopic systems, with the Landau School, and Vitaly Ginzburg in particular, 
having played a prominent pioneering role in interpreting and guiding experimental work \cite{Kapitza1938,Allen1938,Landau1941,Ginzburg2003}. 
Added to this initial motivation, in the last two decades there has been progressive convergence of atomic 
and optical physics on one side, and the study of strongly correlated systems in condensed matter physics on the other. 

Historically, tackling strongly correlated systems in a precise, quantitative way has been one of the hardest goals of theoretical 
physics, probably comparable in difficulty to the attempts to quantitatively understanding strong nuclear interactions. 
This analogy is not accidental, because condensed matter interactions can also be considered too strong to lead 
to quantitatively reliable insights unless numerical computations, whenever feasible, are performed. 
In contrast, noninteracting states of matter have been clearly understood from first principles, as there are two 
classes of particles according to quantum indistinguishability. However, with the singular exception of 
Bose-Einstein condensation of an ideal gas, noninteracting physics cannot obviously capture the dynamics of 
correlated systems and, in particular, the onset of phase transitions. 

Atomic physics allows us to bridge the gap between the noninteracting description and the 
real, strongly interacting case because, given the diluteness of atomic systems, interactions are 
naturally weaker and can then often be treated as perturbations of a noninteracting state. 
Therefore, due to the development of new trapping and cooling techniques, a consistent part of 
atomic physics has now become a sort of dilute condensed matter physics. 
This promises to offer more controllable environments to study prototypical dynamics seen in
less manageable higher-density condensed matter systems. In this regard, a nice experimental surprise 
has been the achievement of the tunability of interatomic interactions among fermions in a continuous way 
and with minimal losses via Feshbach resonances. 

The diluteness of atomic systems has also two further positive features related to enhanced spatial and temporal resolutions. 
The spatial extent and the typical De Broglie wavelength of these systems are larger than in usual condensed matter systems, allowing 
direct imaging of quantum phenomena with optical microscopy resolution, avoiding the complications of submicrometer microscopy. 
Also, the weakness of the interatomic interactions implies slower timescales, and the consequent slower response to external perturbations 
allows the tracking of detailed dynamics. This is in line with the current focus on understanding the dynamics of complex quantum systems, rather 
than their more static description imprinted in the spectral properties of microscopic observables. Probably the best manifestation of the advantages 
of using ultracold atoms is evidenced in the production of quantized vortices \cite{Fetter2001,Fetter2009}, which took many decades of hard and smart 
work on superfluid ${}^4$He \cite{Hall1956,Hess1967,Packard1969,Packard1972}. Vortices were instead produced in copious quantities in atomic 
Bose-Einstein condensates \cite{Aboshaeer2001} to study their spatial correlations in the bulk and the formation of Abrikosov lattices \cite{Abrikosov1957}.
The detailed dynamics of vortex formation \cite{Raman2001} and decay \cite{Aboshaeer2002} have also been investigated. 

Thus, ultracold atomic physics fills the gap between ideal, noninteracting quantum degenerate Bose and Fermi 
gases and their actual counterpart in condensed matter systems like liquid $^4$He and electrons in superconducting 
materials \cite{Pethick2002,Pitaevskii2003}. Degenerate Fermi gases, unlike their bosonic counterparts for which Bose-Einstein 
condensation was achieved in 1995  \cite{Wieman1995,Davis1995,Bradley1995}, have been explored only since 2000. 
Noninteracting, purely quantum mechanical features of dilute Fermi gases were first observed in the degenerate regime, such as 
Pauli blocking \cite{DeMarco1999} and Fermi pressure \cite{Truscott2001,Schreck2001b}, while phenomena involving their interacting 
nature are expected when fermions become highly degenerate. 

Important studies of strongly interacting degenerate Fermi gases were initially reported for Fermi-Bose mixtures 
\cite{Goldwin2001,Modugno2001,Modugno2002,Roati2002}, and for two-component Fermi gases \cite{Ohara2002}. 
In particular, BCS-based models predict a superfluid phase based on Cooper pairing already invoked for the understanding 
of low-temperature superconductivity and superfluidity of $^3$He \cite{Stoof1996}. 
We do not aim here to discuss the interesting physics that has emerged from the study of conventional superfluidity 
in Fermi gases, especially the crossover from tight Fermi pairs (so-called 'molecular' BEC) to the BCS pairing, because   
excellent reviews on this subject are available (see in particular \cite{Chen2005,Giorgini2008,Ketterle2008}).

As usual in real life, there are also various drawbacks to pursuing this research avenue. More specifically, apart from the experimental 
complications of the apparatuses necessary to create these 'artificial' states of matter, correlated states at the atomic level are metastable due 
to various heating and loss sources, including molecular recombination. This limits their usefulness in understanding long-time behavior, such 
as persistent currents, or the precision in measuring crucial parameters, as the critical velocity for the onset of superfluidity \cite{Raman1999,Onofrio2000a}, 
a parameter instead measured with higher precision in liquid helium. From the specific perspective of this review, the greatest 
limitation is the fact that, due to their very dilute nature so advantageous from many standpoints, atomic systems require extremely 
low temperatures to manifest correlations of a quantum nature. While this has provided a strong stimulus to develop innovative 
cooling techniques, it is generally felt that we are reaching a stagnation stage in the achievement of low temperatures. 
This could potentially compromise the exploration of a rich landscape of phase transitions not necessarily driven by a classical 
and macroscopic parameter such as temperature, and which instead occur once the thermal effects are made negligible (so-called quantum phase transitions). 
This issue seems particularly aggravated for fermions, which could allow exploration of model Hamiltonians relevant to understanding 
high-temperature superconductivity, one of the biggest open problems of condensed matter physics today. 
Two main areas explored in atomic physics in this regard are exotic superfluid states, going beyond the standard BCS approach, and the 
study of the Fermi-Hubbard model on optical lattices. Successful experimental studies in both these areas and the possible 
observations of novel quantum phases are conditional on achieving deeper Fermi degeneracy.
 
This review is aimed at summarizing the current status of fermion cooling, with unconventional superfluidity in 
mind as a concrete example of important new physics achievable if fermions are cooled to lower temperatures. 
After a discussion of the quantum degeneracy required to achieve this, we discuss the fundamental limitations 
that occur in adapting to Fermi gases a cooling technique so successful in the case of bosons, evaporative cooling, to Fermi gases, together 
with ways to circumvent these limitations. The most widespread cooling technique, sympathetic cooling through contact with an evaporatively 
cooled Bose gas, is then analyzed in detail for the specific species used in the laboratories. 
This requires a comprehensive discussion of all the experiments reporting observations on quantum degenerate Fermi gases. 
Having identified the basic factors for the limitations on cooling Fermi gases, we then describe various techniques aimed at 
mitigating these limitations, describing their experimental implementations in more detail. A succinct review of proposals for 
cooling optical lattices complements the more detailed description of cooling in single traps, and the interested reader is directed to an excellent review 
already dedicated to optical lattices \cite{McKayRev2011}. The study of phase transitions in general requires the assessment of the temperature of 
the sample, either to ensure that it is low enough to allow the description in terms of quantum phase transitions, or to precisely determine the 
phase diagram in the case of thermally driven, classical phase transitions. This leads to the description of ongoing efforts toward precision 
thermometry of Fermi gases, which together with general remarks in the conclusions completes this review. 

\section{\bf{Searching for exotic superfluid states}}

In the conventional approach to electronic superconductivity, a new ground state is achieved at low temperature if there is an 
effective attractive interaction between the electrons resulting from phonon coupling. This results in a rearrangement of the Fermi sea, the natural 
ground state of a noninteracting Fermi gas, into pairs of electrons with opposite spins and  momenta, the so-called Cooper pairing, which creates 
effective bosons via correlations in momentum space \cite{Cooper1956}. The creation of an energy gap from the Fermi sea protects the pair 
from the usual energy dissipation characteristic of Ohmic conductivity in the normal phase. The corresponding mean field treatment of this approach 
has led to a quantitative, first-principle model of superconductivity, the BCS theory \cite{Bardeen1957}, which contains the Ginzburg-Landau 
model previously developed \cite{Ginzburg1950} as an effective description of the Cooper pairing valid around the critical temperature for 
superconductivity, as shown by Gor'kov \cite{Gorkov1959}. Relevant insights on superconductivity are gained by understanding how it can be 
destroyed, and this occurs by either increasing the temperature above a critical value and/or increasing the external magnetic field. 
Both the critical temperature and the critical magnetic field are characteristic of the substance, the concrete geometry of the sample, and the presence 
of other factors such as external pressure.  In particular, superconductors in strong external magnetic fields 
can manifest a separation of the Fermi surfaces of electrons with opposite magnetic moments. This leads to the Chandrashekar-Clogston limit 
\cite{Chandrashekar1962,Clogston1962}, an upper bound to the ratio between the critical values of the magnetic field and the critical temperature 
for the survival of the superconducting state. Above such a ratio, there is enough magnetic energy to spin-flip the electron antiparallel to 
the magnetic field, thereby destroying Cooper pairs. 

However, it was realized independently by Larkin and Ovchinnikov \cite{Larkin1964} and Fulde 
and Ferrel \cite{Fulde1964} that close to the transition line delimited by the Chandrashekar-Clogston limit, there is the possibility of creating a new 
superconducting state in which spin polarization is accompanied by uncompensated electron momenta. The two mismatched Fermi surfaces can overlap, and 
thus pairing becomes spatially dependent, supporting an inhomogeneous form of superfluidity. The order parameter is still the energy gap, as in the usual 
Cooper pairing, but the gap becomes spatially dependent, with the spatial variability dictated by the De Broglie wavelength associated 
with the difference in the momenta of the two electrons. Such a state is named, from its proponents, the LOFF (or FFLO) state. 
This is an example of a series of alternative pairing mechanisms proposed, including Sarma superfluidity \cite{Sarma1963,Wu2003,Pao2006}, deformed 
Fermi-sphere superconductivity \cite{Muther2002,Sedrakian2005,Sedrakian2006}, and breached pairing superconductivity \cite{Liu2003}. 

Efforts to identify a LOFF state in superconductors have not been conclusive so far. A phase transition in the presence of high magnetic fields has  
been evidenced in low-dimensionality heavy fermions \cite{Gloos1993,Radovan2003,Kakuyanagi2005} and in organic superconductors 
\cite{Dupuis1995}, and tentatively interpreted as a morphing into a LOFF state. However, orbital pair-breaking effects are a competitive source with respect to Pauli 
paramagnetism, from which spin polarization arises. 

Ultracold atomic gases, in principle, provide a cleaner setting in which the effect can be evidenced, and 
unbalanced spin states in ultracold trapped samples of ${}^6$Li have been reported since 2006 \cite{Zwierlein2006,Partridge2006}, including claims for 
the existence of a LOFF phase \cite{Liao2010}. An estimate of the temperature at which LOFF states become the absolute ground state in 
three-dimensional superconductors was given in \cite{Matsuo1998}, with the value of $T = 0.075 T_{BCS}$. Considering the strong dependence of 
$T_{BCS}$ on the fermion density $n$ via the Fermi wave vector $k_F$ and their elastic scattering length $a_s$, $T_{BCS} \simeq T_F \exp[-\pi/(2k_F|a_s|)]$ 
\cite{Gorkov1961,Stoof1996},  in the more optimistic case we would need the Fermi degeneracy factor factor $T/T_F \leq 10^{-3}-10^{-2}$ to observe LOFF superfluidity, 
well below the values achieved in current experimental apparatuses. Here, we have introduced the Fermi temperature $T_F$, which is related to the Fermi energy 
via the Boltzmann constant $k_B$, $E_F=k_B T_F$, the latter being defined for a harmonically trapped gas as 
$E_F=\hbar \omega_F (6N_\mathrm{f}^{1/3}$, where $\omega$ the geometric mean of the trapping angular frequencies along the three directions, and 
$N_\mathrm{f}$ is the number of fermions. Higher values of $T/T_F$ and a broader stability region for a LOFF state are expected for 
lower-dimension systems \cite{Liu2007,Yanase2009}, but direct cooling in lower dimensions may be problematic, and morphing from a 
three-dimensional trap into a one-dimensional one generally results in some heating. The understanding of the limitations of cooling Fermi gases to a 
deeper degeneracy and the search for novel cooling techniques are therefore key elements for searching superconducting states other than those due to Cooper pairing. 

\section{\bf{Generalities of fermion cooling}}

Cooling fermions toward quantum degeneracy has capitalized on the success of cooling bosons to achieve Bose-Einstein condensation. 
For the latter, although purely optical cooling techniques have been pursued from the very beginning, the winning cooling strategy 
has been the use of forced evaporative cooling, a technique more adequate than optical ones in the high-density limit. 
Evaporative cooling relies on fast thermalization after selective removal of more energetic atoms, and therefore requires 
elastic scattering rates large enough compared with competing sources of nonselective loss of atoms: heating 
mechanisms, inelastic scattering, three-body collisions, spin-exchange collisions among others. 
It is natural to consider extending forced evaporative cooling to fermions, but this has a significant limitation 
whenever identical fermions in a single spin state are involved. Indeed, due to the requirement for a totally 
antisymmetric wavefunction, two-body elastic scattering of identical spin-polarized fermions cannot occur 
in an $s$-wave, because in this case the angular part of the wavefunction is as symmetric as the spin part. 
Elastic scattering with higher angular momentum is progressively suppressed at low temperatures, generating  
a bottleneck for thermalization. Three strategies have been imagined to circumvent this issue. 

One possibility is to convert part of the atomic cloud  into another quantum state and proceed with simultaneous evaporative cooling 
of the two distinguishable states, so-called dual evaporative cooling. While intrastate evaporative cooling is still Pauli-suppressed, interstate 
evaporative cooling proceeds unaffected, obviously reducing the number of fermions. 
This strategy is especially suitable for forced evaporative cooling in optical traps, where many hyperfine states are 
trapped simultaneously, and it is progressively advantageous with increasing the hyperfine quantum number $I$, because  
the fraction of states available for $s$-wave elastic scattering with respect to the total available is $2I/(2I+1)$. 
For states with $I=1/2$, this means only $50 \%$, but in the case of $I=5/2$ (for ${}^{173}$Yb, for instance) this means 
that $5/6$ of the possible combinations proceed with evaporative cooling unaffected by the Pauli principle.  
Limits to the minimum temperature achievable by means of dual evaporative cooling have been discussed, including Pauli blocking, 
leading to an estimate of $(T/T_F)_{\mathrm{min}} \simeq 0.3$ \cite{Crescimanno2000,Holland2000}.

Another possibility, also called sympathetic cooling, is to use a bosonic species to indirectly cool the fermions via thermal contact. 
Due to the historical evolution of this subfield, with degenerate bosons predating their fermionic counterpart by five years, and also 
due to the intrinsic interest in studying Bose-Fermi mixtures as natural counterpart to the ${}^3$He-${}^4$He superfluid mixture in the 
liquid state, this is by far the most adopted cooling scheme. The complications of trapping two  different species as happens in various 
mixtures can be offset by the fact that the number of fermions should, in principle, be conserved. 
We mainly focus on this technique in the following considerations.

The third possibility, still under extensive development, is to use atoms in which the long-range atomic forces prevail
over the zero-range pseudopotentials implicit in the mean-field description via a Gross-Pitaevskii equation for bosons and 
analogous Thomas-Fermi approximations for fermions \cite{Baranov2008,Martin2016}. Long-range forces are able to sustain elastic 
scattering with an arbitrary angular momentum at any temperature, and hence the Pauli-blocking bottleneck should be avoided for a single species, 
single state fermions. Very few fermionic species have dipolar forces naturally dominating over the zero-range pseudopotential 
interaction, and so far attention has been focused on dysprosium  \cite{Lu2012,Burdick2015,Burdick2016} and erbium 
\cite{Aikawa2014a,Aikawa2014b}.  In particular, a ${}^{161}$Dy-${}^{162}$Dy Fermi-Bose degenerate mixture has been 
realized, reaching the Fermi degeneracy factor $T/T_F=0.2$ \cite{Lu2012}. Even more intriguing, forced evaporative cooling of spin-polarized 
${}^{162}$Dy alone allowed reaching $T/T_F=0.7$, a result attributed to the presence of long-range scattering for a polar gas.
Larger elastic collision rates have also associated larger inelastic dipolar collisions in metastable magnetic states; however, it has been 
predicted \cite{Pasquiou2010} and experimentally verified \cite{Burdick2015} that quantum statistics give a suppression mechanism in the case 
of fermionic species. The presence of favorable elastic scattering in the Fermi degenerate regime has been confirmed by using ${}^{167}$Er atoms, 
bringing $6.4 \times 10^4$ fermions at $T/T_F=0.2$ \cite{Aikawa2014a} and studying the anisotropic relaxation dynamics of the cloud \cite{Aikawa2014b}, 
obtaining the lowest value of $T/T_F=0.11\pm 0.01$. The Fermi degeneracy factors for cooling dipolar gases achieved so far are comparable 
to the ones achieved in $s$-wave sympathetic cooling of Fermi-Bose mixtures. However, the presence of anisotropies, the need for a strong magnetic field precluding 
Feshbach-like control of the elastic cross section, and complications in the precooling stage due to the complex spectroscopy of dysprosium and erbium atoms 
may preclude the broad diffusion of this approach. 

\begin{figure*}[t]
\begin{center}
\includegraphics[width=0.33\textwidth]{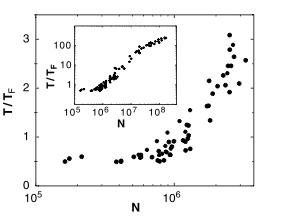}
\includegraphics[width=0.33\textwidth]{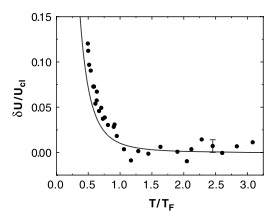}
\includegraphics[width=0.30\textwidth]{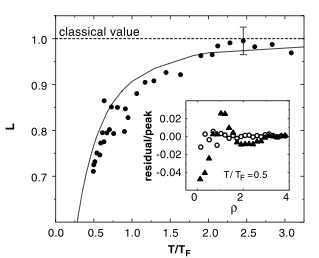}
\caption{Evidence of quantum degeneracy of ${}^{40}$K in the JILA experiment. 
From left to right, $T/T_F$ versus number of fermions, showing a flattening 
region at the lowest value achieved, $T/T_F \simeq 0.5$, an excess of internal energy 
versus $T/T_F$, and deviations from Gaussianity of the momentum distribution
at low $T/T_F$, expressed through a shape indicator with a unit value for a Gaussian 
shape. The inset shows the greater discrepancies arising from a Gaussian fit (solid triangles)
compared with a non-Gaussian fit (white dots) (reproduced from \cite{DeMarco1999}).}
\label{}
\end{center}
\end{figure*}

\section{\bf{Dual evaporative cooling of fermions}}

The first realization of a Fermi degenerate gas in the dilute atomic physics setting was achieved by Brian DeMarco and Deborah Jin from 
JILA and the University of Colorado \cite{DeMarco1999}. Atoms were trapped in a double magneto-optical trap (MOT) apparatus in 
which the first MOT was used to collect large amounts of atoms from a potassium dispenser enriched in the fermionic isotope ${}^{40}$K. 
The atoms were then delivered to the high-vacuum MOT through the radiation pressure exerted by laser pulses. 
In the second MOT, the atoms were cooled to their Doppler limit of 150 $\mu$K, and loaded into a Ioffe-Pritchard 
magnetic trap, where the trapping lifetime was measured to be 300 s. This is long enough to allow for forced dual evaporative 
cooling. Evaporative cooling requires states that are stable, at least on the timescale over which experiments are performed, against 
$m_F$-changing collisions at low temperatures, and in this case the choice was made for $|F=9/2, m_F=9/2 \rangle$ and 
$|F=9/2, m_F=7/2 \rangle$. Selective removal was obtained by applying microwave transitions with two distinct frequencies tuned 
to an untrapped spin state of the F=7/2 manifold. The magnetic field was varied to keep the fraction of removed atoms in the two states 
within 5 $\%$ of each other.  Absorption imaging of the cloud after release allowed measuring the temperature and the number of atoms, and 
then mapping the evaporation trajectory as can be seen in Fig. 1. The Fermi temperature was evaluated by measuring the number of fermions and radial and 
axial trapping frequencies. 

It is evident from Fig. 1 that while the evaporation trajectory is steep for $T/T_F \geq 1$, it flattens when the gas enters quantum degeneracy, 
reaching a constant value $T/T_F \simeq 0.5$. The experimenters did not observe any concomitant change in atom losses or heating rate, and 
the decrease in efficiency persisted even through various changes in the trapping parameters. In particular, changing the number of atoms in the 
range between $3.5 \times 10^5$ and $1.2 \times 10^6$ and the geometric mean frequency in the range from 127 Hz to 323 Hz, corresponding to changes 
in the Fermi energy in the range from 0.36 to 1.0 $\mu$K, confirmed that evaporative cooling always stalled at an absolute temperature of about $0.5 T_F$. 
Fermi pressure, limiting the shrinking of the cloud and therefore the increase in the density crucial for increasing the elastic scattering 
rate, and Pauli blocking were put forth as the explanation of the observed phenomenon. 
In addition, the total energy was measured from the second moment of absorption images of the expanded clouds, under conditions such that interactions 
could be neglected due to the low-density regime. In coincidence with $T \simeq 0.5 T_F$, a sharp increase in the total energy was observed with respect 
to that of the corresponding classical gas. The excess of energy was also manifested in a non-Gaussian momentum distribution. 

Another group, at Duke University, adopted an all-optical trapping and cooling approach to two 
hyperfine states of ${}^6$Li. Because magnetically trappable states in ${}^6$Li, unlike their ${}^{40}$K counterparts,
are vulnerable to spin-exchange collisions and dipolar decays, the adopted strategy consisted in avoiding the magnetic 
trapping stage by loading the atoms directly from a MOT into a single-beam optical dipole trap (ODT) created by a strongly 
focused (47 $\mu$m waist, Rayleigh range of 660 $\mu$m), high power CO${}_2$ laser at the wavelength of 10.6 $\mu$m. 
The corresponding initial trapping frequencies for the radial and axial confinement were respectively measured to be 6.6 KHz and 340 Hz. 
The use of a CO${}_2$ laser results in an extremely low residual Rayleigh scattering rate due to the separation of its wavelength from 
the dominant atomic transition (D${}_2$ line at 671 nm). For the peak intensity of 1.0 MW cm${}^{-2}$, an average of two photons per hour 
are scattered, corresponding to a recoil heating rate of 16 pK s${}^{-1}$. This may be crucial in the deep degenerate 
regime since, as temperature  decreases, small heating rates potentially enormously harm the 
temperature of the cloud due to the sensitive dependence of the specific heat on temperature.

A crucial advantage of an optical dipole trap is that the lower densities expected during 
evaporative cooling due to the decreasing laser power can be compensated by a bias magnetic field 
tuned to a Feshbach resonance to maximize the elastic scattering length.
From this perspective, the use of hyperfine states $|F=1/2, m_F=\pm 1/2 \rangle$ is favorable because 
a Feshbach resonance around 850 G was predicted \cite{Houbiers1997}, having at the same time 
a zero scattering length in a zero magnetic field. Full control of the scattering rate and runaway 
evaporation may therefore be achieved by means of the bias magnetic field. 

The Duke group has achieved the Fermi degeneracy $T/T_F$=0.55 with $3 \times 10^5$ atoms \cite{Granade2002}. In subsequent experiments, 
the group succeeded in producing a rapid expansion of the cloud in the radial direction of confinement, with a nearly 
constant size in the axial direction \cite{Ohara2002}, a feature not shared by the experiment performed with ${}^{40}$K at Boulder. 
Data were interpreted with two alternative models, as a collisionless superfluid and in terms of collisional hydrodynamics, and 
arguments of plausibility in favor of the first model were given. 

After these two groups, various laboratories have used dual evaporative cooling for a variety of demonstrations 
and experiments, and we briefly mention some of them. Trapping ${}^6$Li in an ODT with a single Yb:YAG laser beam at the wavelength of 1030 nm with proper 
tuning of the magnetic field near a Feshbach resonance allowed creating a Bose-Einstein condensate of ${}^{6}$Li${}_2$ molecules \cite{Jochim2003,Fuchs2007}.
Fermionic lithium cooled with dual evaporation and trapped in the antinodes of a standing wave has allowed the 
realization of a two-dimensional degenerate gas with $T/T_F=0.10 \pm 0.03$ \cite{Martiyanov2010a,Martiyanov2010b}, 
and radiofrequency spectroscopy of a two-dimensional gas of ${}^{40}$K has also been implemented to measure the interaction energy 
\cite{Frohlich2011}. 

Other groups have used dual evaporative cooling to study ultracold collisional properties of three-state mixtures of 
${}^6$Li \cite{Ottenstein2008,Huckans2009}, and to create ${}^6$Li${}_2$ molecules through $p$-wave Feshbach resonances \cite{Inada2008}. 
Besides lithium and potassium, dual evaporative cooling has also been used to bring ${}^{87}$Sr to Fermi degeneracy, with a remarkable 
$T/T_F =0.26$ \cite{DeSalvo2010}. 

Dual evaporative cooling is also currently used as the last stage after sympathetic cooling, especially if the latter occurred in 
a magnetic trap. The Bose gas is removed and the spin-polarized Fermi gas is loaded into an optical dipole trap. This is 
then followed by the preparation of a two spin state mixture using a radiofrequency pulse and intentional decoherence.
The price to proceed with this further stage of cooling is that the number of fermions in each spin component, for an equally 
balanced mixture, is half the total number,  thereby increasing the Fermi degeneracy parameter $T/T_F$. Another possibility 
is to use a Feshbach resonance to produce a molecular BEC of fermions, and evaporatively cool the effective bosonic sample, without incurring 
the Pauli suppression of elastic scattering. At the end of the process, the magnetic field can be tuned to a value corresponding to 
loose pairing in space, on the BCS side of the Feshbach resonance. Furthermore, in this case, the presence of losses and deviations from adiabaticity in this 
magnetic field sweep procedure may easily increase the Fermi degeneracy parameter.  

\begin{figure*}[t]
\begin{center}
\includegraphics[width=0.30\textwidth]{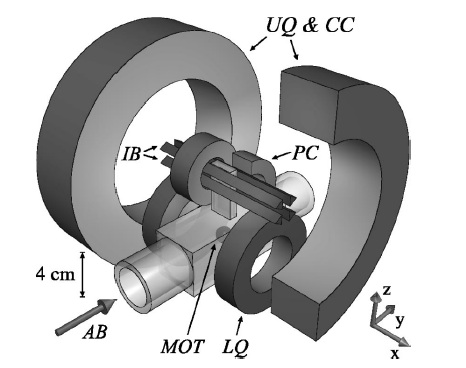}
\includegraphics[width=0.69\textwidth]{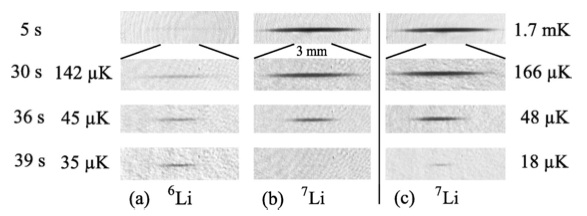}
\caption{(left) Apparatus for trapping and cooling bosonic and fermionic lithium 
isotopes coming from a slow atomic beam (AB), initially trapped and cooled 
in a magneto-optical trap (MOT), then transferred into an Ioffe-Pritchard (IP) trap. 
Also shown are the coils used as an atom magnetic elevator, lower quadrupole (LQ) 
and upper quadrupole (UQ), and the compensation and pinch coils used for evaporative 
cooling (CC and PC). (right) Absorption images of the two clouds during (b and c) sympathetic 
cooling, and (d) single-species (Bose gas) evaporation (reproduced from \cite{Schreck2001a}).}
\label{}
\end{center}
\end{figure*}

\section{\bf{Fermi-Bose mixtures}}

An alternative to dual evaporative cooling of different states of the same atomic Fermi species is indirect cooling via thermal 
contact with a reservoir of Bose atoms. This does not suffer from the limitations of evaporative cooling due to the quantum statistics discussed above. 
First demonstrated in the atomic physics context in \cite{Myatt1997}, sympathetic cooling has generated a subfield in itself, the study of 
ultracold atomic and molecular mixtures, with peculiar phenomena such as induced collapse and demixing depending on the sign and amplitude of the interspecies elastic scattering length \cite{Molmer1998}. 

On the negative side, sympathetic cooling requires more complicated apparatuses, because two species need to be simultaneously trapped and cooled. 
Sympathetic cooling proceeds efficiently as long as the heat capacity of the Bose gas is large or at least comparable to the heat capacity of the Fermi gas. 
In the classical limit this is mainly a statement on the dominance of the number of bosons compared to the number of fermions, because the specific heats 
are the same in the Dulong-Petit, nondegenerate limit. Below the critical temperature for Bose-Einstein condensation, 

$$T_c=\zeta(3)^{-1/3}\hbar \omega_f N_\mathrm{b}^{1/3}/k_B$$
and the Fermi temperature  

$$T_F=6^{1/3}\hbar \omega_f N_\mathrm{f}^{1/3}/k_B,$$
both defined in terms of the mean angular trap frequencies $\omega_b=(\omega_{b}^{(x)} \omega_{b}^{(y)} \omega_{b}^{(z)})^{1/3}$ and 
$\omega_f=(\omega_{f}^{(x)} \omega_{f}^{(y)} \omega_{f}^{(z)})^{1/3}$, the boson and fermion heat capacities approach zero with the respective 
temperature dependences $T^3$ and $T$. If $\omega_f=\omega_b$, the boson heat capacity becomes smaller than the fermion one below 
$T/T_F\simeq 0.3$, strongly affecting the  efficiency of sympathetic cooling for smaller $T/T_F$, consistently with the limit value 
of $T/T_F\simeq 0.25$ reported in the initial experiments \cite{Schreck2001a,Truscott2001,Schreck2001b} where $^7$Li-${}^{6}$Li mixtures were used. 
While the presence of the Bose gas appears to be irrelevant in the latter stage of cooling in the degenerate regime, it actually worsen things, depending 
on the magnitude of its associated heating rate with respect to the cooling rate. Indeed, the large sensitivity of the Bose gas to a small release of 
heat due to the cubic scaling of the specific heat at low temperature increases its temperature considerably. 
This is potentially detrimental to the Fermi gas, in spite of its dominant heat capacity in the same temperature range, and can result in an  
efficient but undesired form of sympathetic heating. Removal of the Bose gas in the latter stage of the cooling protocol is a possible solution 
to this issue, followed by dual evaporation of the Fermi gas. 

In what follows, we limit our analysis to the aspects of Bose-Fermi mixtures relevant to sympathetic cooling.  
The analysis proceeds with a description of all the mixtures, approximately in historical order, and culminates with 
a summary of the minimum $T/T_F$ achieved. We focus only on Fermi-Bose mixtures in which the fermions have been brought to the quantum 
degenerate regime, {\it i.e.} $T/T_F <1$, and for brevity do not consider more sophisticated setups involving more than two atomic species, such 
as Fermi-Fermi-Bose mixtures. Even so, and considering the accelerating growth of experiments on Fermi-Bose mixtures, we apologize in 
advance if some experiments may have been overlooked or not sufficiently highlighted as desired by the involved groups.

\subsection{\bf{${}^6$Li-${}^{7}$Li}}

This mixture has been trapped and cooled in the laboratories of the \'Ecole Normale Sup\'erieure in Paris \cite{Schreck2001a,Schreck2001b} and 
Rice University \cite{Truscott2001}. Among the advantages of using this mixture, the experimental set up is relatively easy because  
the light beams necessary for magneto-optical trapping differ in wavelength only by isotopic shifts, well within the range of tunability 
by acousto-optical and electro-optical modulators. This allows the use of a common laser source, minimizing the effort for daily maintenance. 

Efficient sympathetic cooling of ${}^6$Li via ${}^7$Li was based on predictions of a favorable elastic scattering length 
\cite{VanAbeleen1997}. However, the presence of a limitation to the maximum number of condensed bosons for a specific ${}^7$Li state
makes the bosonic reservoir less appealing. Fermionic ${}^6$Li was also regarded as a promising candidate due to the intrinsically large and 
negative elastic scattering length, which was expected to facilitate BCS pairing at moderately low temperatures \cite{Combescot1999}. 

Many peculiar elements of the Bose-Fermi mixtures were first addressed in these two experiments, providing milestones for subsequent 
investigations by other groups. In particular, the advantages of using the Bose component not only as a coolant but also as a 
precision thermometer, the need for checking the thermal equilibrium between fermions and bosons at all stages  and, most important to 
our discussion, the observed decreased efficiency of sympathetic cooling in the degenerate regime. This last issue results from the 
combined effect of the continuous depletion of ${}^7$Li atoms during evaporative cooling and the drop in their specific
heat below the BEC phase transition. A numerical estimate of the minimum $T/T_F$ achievable was consistent with observations. 
Also, the need for maximal spatial overlap of the two species was emphasized. 

\begin{figure}[b]
\begin{center}
\includegraphics[width=0.95\columnwidth]{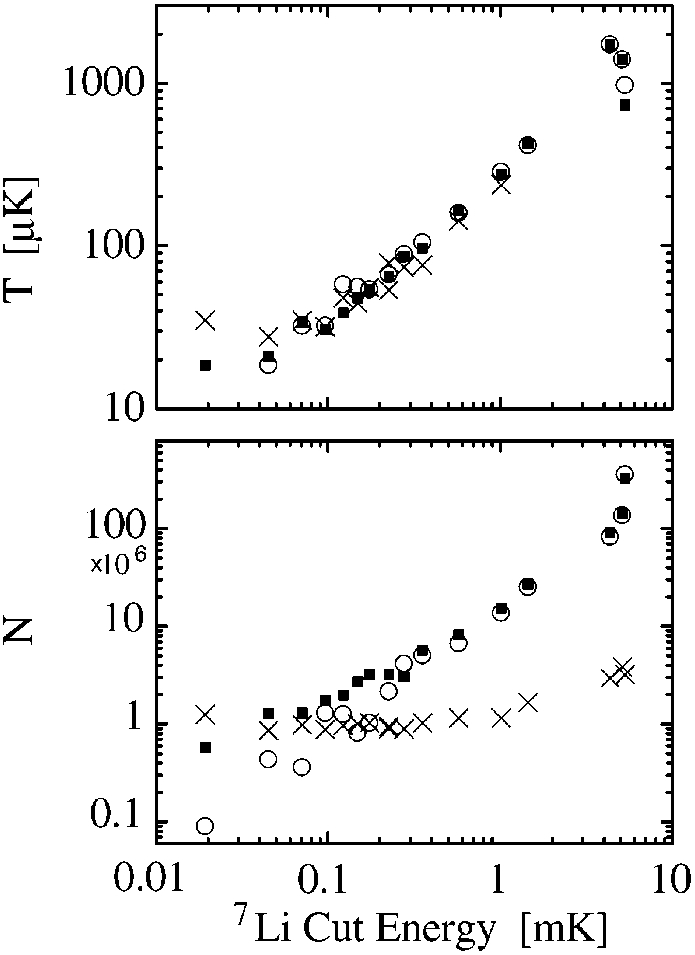}
\caption{Temperatures (above) and number of atoms (below) versus the threshold energy 
for evaporative cooling of ${}^{7}$Li (white circles) in the presence of ${}^6$Li (crosses). 
The fermionic component follows the bosonic component until the last stage at the lowest 
energies, for which there is evidence of an out-of-equilibrium state. The bosonic 
component is unaffected by the fermionic one, as shown (black squares) by evaporative cooling of 
${}^{7}$Li alone (reproduced from \cite{Schreck2001a}).}
\label{}
\end{center}
\end{figure}

The Paris group has explored an ingenious way to increase quantum degeneracy by stiffer confinement in a compact Ioffe-Pritchard trap, 
achieved by transporting the atoms initially cooled in a MOT into a small appendage of the glass cell. With operating currents of 700 A 
for the Ioffe bars and 500 A for two pinch coils, the confinement results in trap frequencies of 2.57 KHz in the radial direction and 
117 Hz in the axial direction. These large values of the trapping frequencies resulted in an unusually large value of the corresponding Fermi temperature. 
Evidence of a decoupling between the temperatures of ${}^6$Li and ${}^7$Li was reported at a temperature of 35 $\mu$K, at which the number of atoms 
in the two species and the heat capacities were approximately equal. The group tried to increase the cooling capability by removing fermions, but this resulted 
in a number of atoms below the sensitivity threshold of the imaging system. In the first report, the group obtained $1.3 \times 10^5$ ${}^6$Li 
at a temperature of $9\pm 3 \mu$K, corresponding to the degeneracy factor $T/T_F=2.2 \pm 0.8$ \cite{Schreck2001a}. 
In their second report \cite{Schreck2001b}, the ENS group succeeded in obtaining a Fermi degeneracy factor similar to the one 
reported earlier by the Rice group, of the order of $T/T_F \simeq 0.3$. Two different hyperfine states for bosonic lithium were investigated:
${}^7$Li prepared in the $|2 2\rangle$ hyperfine state, for which the negative interspecies scattering length (-1.4 nm) results in  a collapse 
of the condensate above a threshold in the number of atoms of about 1,400, and a more favourable hyperfine state, $|1 -1 \rangle$, in 
which this limitation does not occur due to a positive (0.27 nm) elastic scattering length, although this results in an elastic cross-section 
smaller by a factor of about 27 compared with the one in the $|2 2 \rangle$ hyperfine state. To compensate for the small scattering length, 
${}^6$Li was used as mediator for the interspecies thermalization of ${}^7$Li. 

The Paris group has recently reported studies on a mixture of Bose and Fermi superfluids \cite{Ferrier2014} for different values of the 
elastic scattering length among fermions and a study of the relative critical velocity for superfluidity in this mixture \cite{Delehaye2015}, 
reporting the minimum Fermi degeneracy $T/T_F \simeq 0.03$. 

\subsection{\bf{${}^6$Li-${}^{23}$Na}}

This mixture, extensively used by the ultracold atoms group at the Massachusetts Institute of Technology, was the first consisting 
of two different atomic species. The use of ${}^{23}$Na as a coolant was felt to be advantageous compared with ${}^7$Li because of the limitations of the 
latter described in the preceding subsection. In addition, fast interspecies thermalization was also observed, with low rates for intraspecies and interspecies 
inelastic collisions allowing lifetimes in excess of 10 s. A ${}^6$Li beam line was inserted with minimal changes compared with one of 
the existing apparatuses already used to produce large ${}^{23}$Na condensates \cite{Hadzibabic2002}. 
This resulted in the choice of a two-species oven with separate chambers, due to a three-orders-of-magnitude difference in the vapor pressure of ${}^6$Li and ${}^{23}$Na.
A common Zeeman slower, due to the smaller mass of lithium, allows slowing lithium without interfering with the slowing efficiency of sodium.
A rather complex optical pumping procedure was used to prepare the mixture in states protected against spin-exchange collisions while still 
magnetically trappable after transfer from the MOT stage.  In a later report, the MIT group reported a record number of ${}^6$Li in the degenerate 
regime with up to $7 \times 10^7$ atoms and the Fermi degeneracy factor of 0.5, reduced to $3 \times 10^7$ atoms at the Fermi degeneracy factor 
in the $0.05 \div 0.2$ range \cite{Hadzibabic2003}. This has allowed the transfer of large clouds of ${}^6$Li atoms into an 
optical dipole trap, preparing the atoms in two distinct states.  This was followed by dual evaporative cooling in the presence of a constant magnetic
field tuned for Feshbach resonances, leading to various results like the first observation of quantized vortices in a lattice pattern in the BCS regime 
\cite{Zwierlein2005,Zwierlein2006,Ketterle2008}. 

\subsection{\bf{${}^{40}$K-${}^{87}$Rb}}

This mixture has the practical advantage of the proximity of the main transition wavelengths (767 nm and 780 nm),  
for ${}^{40}$K and ${}^{87}$Rb, which allows low-cost and low-power diode lasers to be used for seeding a common 
high-power tapered semiconductor amplifier chip, considerably simplifying the setup for the double MOT. 
This approach has been first pursued at JILA \cite{Goldwin2001,Goldwin2004} and at LENS in Florence 
\cite{Modugno2001,Modugno2002,Roati2002}. 
A strong interspecies attraction was observed in Florence, leading to instability above a threshold in the particle numbers. 
More specifically, the loss rate of ${}^{40}$K was observed to depend quadratically on the ${}^{87}$Rb density, quantitatively 
explainable by assuming a three-body K-Rb-K recombination rate of $2(1) \times 10^{-27} \mathrm{cm^6 s^{-1}}$.
Although this precluded the full exploitation of ${}^{87}$Rb as a coolant for ${}^{40}$K, the Florence group pointed 
out that the large boson-fermion attractive interaction should imply an effective attractive fermion-fermion interaction, leading 
to a situation quite similar to phonon-mediated interactions in the BCS model of conventional, low-$T_c$ superconductivity.
This configuration was suggested as an alternative to the enhancement of fermion-fermion interactions through the use of Feshbach 
resonances or of strongly confining optical lattices. The JILA group also achieved simultaneous quantum degeneracy at the 
lowest $T/T_F=0.2$,  and measured an interspecies elastic scattering length lower than the one measured by the Florence group. 

Issues of stability and instability due to the interspecies interactions have been systematically addressed in \cite{Ospelkaus2006}, 
where large numbers of fermions and bosons allowed determining the threshold for mean-field-driven collapse of the mixture.
The Florence group also reported preparation of a degenerate Fermi gas confined in a one-dimensional optical lattice 
\cite{Modugno2003}, while a three-dimensional optical lattice of degenerate ${}^{40}$K was realized at ETH in Zurich \cite{Kohl2005}. 
In \cite{Aubin2006}, rapid sympathetic cooling (evaporation time of 6 seconds) was achieved using an atom chip.
This mixture was used to create a gas of polar molecules close to quantum degeneracy ($T/T_F \simeq 2$)   \cite{Ni2008},  
and to demonstrate collective atomic recoil of a degenerate Fermi gas \cite{Wang2011}. For this last experiment, after sympathetic cooling in a 
Ioffe-Pritchard magnetic trap and the subsequent removal of all ${}^{87}$Rb atoms, an adiabatic expansion of the ${}^{40}$K atoms was quickly 
performed, bringing the cloud to a different aspect ratio. Due to the lack of thermalization of the ${}^{40}$K atoms alone, this led to a 
momentum-squeezed state for the cloud, which was then studied under a pump laser pulse with controllable polarization, resulting in multiple wave-mixing processes. 

\subsection{\bf{${}^6$Li-${}^{87}$Rb}}

In theory, this mixture has appealing features from various standpoints. The large electric dipole moment expected 
in LiRb molecules makes this mixture a promising candidate for quantum computers and for tests of time-reversal invariance.
It has been suggested that, within stable alkali species, this mixture is optimal for large trapping frequency ratio, beneficial for the 
reasons discussed in the next section. Also, the large mass ratio should make fermions less vulnerable to Fermi hole heating 
\cite{Timmermans2001}, as discussed in detail in \cite{Cote2005}. However, the large mass ratio also has drawbacks, in particular with regard to the spatial 
overlap due to gravitational sagging, mitigated in a stiff confinement, and the effectiveness of the elastic scattering.

A group at the University of T\"ubingen reported degeneracy in a ${}^6$Li-${}^{87}$Rb mixture in a magnetic trap \cite{Silber2005}. 
Unfortunately, the interspecies elastic scattering length measured via cross thermalization was small, $a=20^{\mbox +9}_{\mbox -8} a_B$ 
(with $a_B$ being the Bohr radius), preventing its use for efficient driving the Fermi gas to the lowest temperature achievable by rubidium atoms. 
The same group identified two Feshbach resonances after transferring the clouds into a crossed optical dipole trap \cite{Deh2008}. 
However, these Feshbach resonances are accompanied by a dramatic increase in the cross section for inelastic collisions, which 
may be considered beneficial for forming heteronuclear molecules, but not for the goal of achieving deeper Fermi degeneracy. 

Further theoretical work by a group at British Columbia allowed the prediction of other resonances based on the measured triplet 
scattering length and Feshbach resonances for one specific pair of hyperfine states \cite{Li2008}. The experimental study of  
six large Feshbach resonances in ${}^{6}$Li-${}^{87}$Rb by the same group also evidenced large inelastic collision rates 
detrimental to the lightest species and resulting in prohibitive losses from the optical dipole trap \cite{Deh2010}. 

On the instrumental side, high-power solid state lasers are available for the relevant atomic transitions of both atoms, leading to a 
standardization of the operation and maintainance of the apparatus. 

Due to the extreme difference in the vapor pressure of Li and Rb, various configurations have been realized for loading the 
mixture to an MOT, with some solutions having been later adopted for other mixtures. 
In T\"ubingen, Rb atoms were collected in an MOT from a dispenser, while Li atoms were delivered via an oven and a Zeeman slower. 
Alternative schemes were tested at Dartmouth \cite{Brown2007} with two independent Zeeman slowers, at University of British Columbia 
with two dispensers \cite{Ladouceur2009}, and at the University of California, Berkeley with a common Zeeman slower fed by a two-reservoir 
effusive oven \cite{Marti2010}, analogously to the ${}^6$Li-${}^{23}$Na case.

\subsection{${}^{3}$He-${}^{4}$He}

A group at the Vrjie University  in Amsterdam has reported simultaneous quantum degeneracy for the ${}^3$He-${}^4$He mixture \cite{McNamara2006}, the 
gaseous counterpart of the well-known ${}^3$He-${}^4$He mixture in the liquid state. The atoms were trapped in their lowest triplet state, which is metastable 
with subnanosecond lifetimes in the absence of external magnetic trapping. The internal energy is about 11 orders of magnitude larger than their 
average thermal energy at the lowest temperatures explored (of the order of 1 $\mu$K), which allows high-efficiency single-atom detection with high spatial 
and temporal resolution using microchannel plate detectors. This allowed studying degenerate gases with high temporal 
resolution, with the statistical analysis used in quantum optics, such as bunching and antibunching phenomena \cite{Jeltes2007}, extended to them, and 
exploring precision metrology of simple atoms \cite{Vassen2012}. A detailed study of the measurement of the temperature of the mixture using the 
dependence of the voltage signal in the microchannel plate detector on the time of flight was presented in \cite{McNamara2006}. 
In the same paper the major drawback of this system was also highlighted: the impossibility of using Feshbach resonances, because the atoms need to 
be in fully stretched magnetic states during the trapping procedure.

\subsection{${}^{171}$Yb-${}^{174}$Yb and ${}^{173}$Yb-${}^{174}$Yb}

Ytterbium, a rare-earth element, has peculiar properties appealing to atom trappers. First, it occurs in a variety of isotopes, of both a  bosonic 
(${}^{172}$Yb, ${}^{174}$Yb, ${}^{176}$Yb) and a fermionic (${}^{171}$Yb and ${}^{173}$Yb) nature, all with natural abundances 
larger than 10$\%$. This allows broad combinatorics of Fermi-Bose mixtures with minimal changes to the MOT and post-MOT 
trapping setups, basically achievable by electro-optic and acousto-optic modulators. Second, a weak intercombination transition 
$1S_0 \rightarrow {}^3P_1$ has a narrow linewidth corresponding to a Doppler temperature of 4.4 $\mu$K, allowing 
efficient precooling and loading into an optical dipole trap with moderate laser power. The absence of a magnetic moment 
in the ground state makes optical trapping a must and is appealing for metrological studies due to the negligible Zeeman shifts. 

A group at Kyoto University has pioneered studies of this element in a variety of configurations, including mixtures with other atomic species, which we 
defer to the next sections, focusing here on ytterbium mixtures. Trapping of ${}^{171}$Yb and ${}^{174}$Yb was reported in a crossed dipole 
trap after attempts to perform evaporative cooling on ${}^{171}$Yb alone \cite{Honda2002}. 
The experiment demonstrated an increase in the phase space density of the Fermi gas when the Bose cloud was also present, indirect 
evidence for effective sympathetic cooling, but no quantum degeneracy was reported for either species. 
In subsequent efforts, a regime of quantum degeneracy was progressively achieved, including evaporative cooling of ${}^{171}$Yb alone, 
both in a 1D optical lattice and in a large-volume optical dipole trap, for which the lowest measured degeneracy parameter was $T/T_F=0.37 \pm 0.06$ 
in the case of ${}^{173}$Yb, which has a favorable elastic scattering length \cite{Takasu2006,Fukuhara2007a}. 

In another report, the achievement of Fermi degeneracy for ${}^{171}$Yb via sympathetic cooling to ${}^{174}$Yb was 
reported, overcoming the small elastic scattering length of ${}^{171}$Yb \cite{Fukuhara2007b,Fukuhara2009a}, and 
collective excitations in the mixture were also explored \cite{Fukuhara2009b}. 

We note that due to the large hyperfine quantum number, ${}^{173}$Yb tends to be efficiently cooled by dual evaporation without the need for a refrigerant, because 
only one out of six channels for elastic scattering is Pauli-frozen due to its large ($I=5/2$) nuclear spin, at variance with the case of ${}^{171}$Yb, which has $I=1/2$.
 
\subsection{${}^{87}$Sr-${}^{84}$Sr}

This mixture is particularly interesting due to the large nuclear spin of ${}^{87}$Sr, I=9/2, corresponding to a 10-fold degererate ground state, 
allowing the investigation of model Hamiltonians based on SU(10) symmetry. 

The Innsbruck group succeeded in producing a single-spin degenerate Fermi gas of ${}^{87}$Sr via sympathetic cooling with the bosonic counterpart, 
${}^{84}$Sr \cite{Tey2010}. The presence of interisotope collisions is beneficial to a reduction in atomic losses, as demonstrated by applying the same 
evaporative cooling procedure used for the mixture to ${}^{87}$Sr alone, with a significant reduction in the number of atoms and Fermi degeneracy factor.

\subsection{${}^{6}$Li-${}^{174}$Yb}

The Kyoto group has more recently focused activity on a mixture made of fermionic Li and bosonic 
Yb, especially ${}^{174}$Yb \cite{Hara2011}. The main interest is in the creation of molecules 
with uncompensated electrons in the ground state, the large electron dipole moment 
being advantageous for spin-lattice quantum simulators. The large mass ratio of 29 allows also 
lithium to be in a deep degenerate regime, having ytterbium nearly or below Bose degeneracy.
\begin{table*}[t]
\begin{center}
\begin{tabular}{|l|c|c|c|c|c|c|c|}
\hline
Fermi-Bose Mixture                                & $T/T_F$            & $N_\mathrm{f}$                      &  $N_\mathrm{b}$                     & $\omega_f/\omega_b$ & Reference & Institution & Year\\
\hline 
 ${}^6$Li-${}^7$Li                & $0.25$                              &            $1.4 \times 10^5$          & $2.2 \times 10^4$    &  $1.08$           & \cite{Truscott2001}  &Rice University   & 2001   \\
 ${}^6$Li-${}^7$Li                & $0.2\pm 0.1$                   &             $4 \times 10^3$              & $10^4$                   &  $1.08$            & \cite{Schreck2001b} & ENS Paris & 2001   \\
 ${}^{40}$K-${}^{87}$Rb       &  $0.30$                             &   $10^4$                                       & $2 \times 10^4$   &    $1.47$           &  \cite{Roati2002}      & LENS Florence & 2002  \\
 ${}^6$Li-${}^{23}$Na          &  $0.05{}^{\mbox +0.03}_{\mbox -0.02}$  &  $3 \times 10^7$  & $6 \times 10^6$   &   $1.94$            &  \cite{Hadzibabic2003}  & MIT & 2003   \\
${}^{40}$K-${}^{87}$Rb       &  $0.20$                             &   $10^4$                                       & $2.5 \times 10^5$   &    $1.47$           &  \cite{Goldwin2004}      & JILA Boulder & 2004  \\
 ${}^{40}$K-${}^{87}$Rb       &  $0.32$                             &  $6 \times 10^5$                         & $4 \times 10^5$    &   $1.47$                 &  \cite{Kohl2005}  & ETH Zurich & 2005   \\
 ${}^{3}$He-${}^{4}$He        &  $0.45$                              &   $10^6$                          &  $10^6$                 &    $1.15$           &  \cite{McNamara2006}   & Vrjie University Amsterdam & 2006 \\
 ${}^{40}$K-${}^{87}$Rb       &  $0.1$                               &   $9 \times 10^5$                       & BDL  &    $1.47$           &  \cite{Ospelkaus2006} & Instit\"ut f\"ur Laserphysik Hamburg & 2006  \\
 ${}^{40}$K-${}^{87}$Yb       &  $0.9$                                &  $2 \times 10^4$                        &  BDL &    $1.47$           &  \cite{Aubin2006}  & University of Toronto & 2007   \\
 ${}^6$Li-${}^{87}$Rb          &   $0.90$                             &  $1.4 \times 10^5$                     & $4 \times 10^6$     &     $2.5$          &  \cite{Deh2008}             & Universit\"at Tubingen & 2008 \\
 ${}^{173}$Yb-${}^{174}$Yb &  $0.3$                                &  $10^4$                                     &  $3 \times 10^4$     &     $1.00$            &   \cite{Fukuhara2009a}  & Kyoto University & 2009\\
 ${}^{6}$Li-${}^{174}$Yb      &  $0.08  \pm 0.01$             &  $2.5 \times 10^4$                     & $1.5 \times 10^4$   &     $3.90$            &   \cite{Hara2011}           &  Kyoto University & 2011  \\
 ${}^{6}$Li-${}^{174}$Yb       &  $0.3$                                &  $1.2 \times 10^4$  & $2.3 \times 10^4$   &     $8.20$            &   \cite{Hansen2011}      & University Washington Seattle & 2011 \\
 ${}^{40}$K-${}^{87}$Rb       &  $0.3$                                &  $2.0 \times 10^6$                     & $10^5$                  &    $1.47$            &   \cite{Wang2011}      & Shanxi University & 2011  \\
 ${}^{40}$K-${}^{23}$Na       &  $ 0.35$                            & $3.0 \times 10^5$                       & $10^6$                 &    $-$              & \cite{Park2012}  &    MIT & 2012 \\
${}^{87}$Sr-${}^{84}$Sr        &  $ 0.30 \pm 0.05$               & $2.0 \times 10^4$                       & $10^5$                 &    $0.98$              & \cite{Tey2010}  &  Universit\"at Innsbruck & 2012 \\
 ${}^6$Li-${}^7$Li                & $0.03$                              &  $2.5 \times 10^5$                      & $2.5 \times 10^4$  &      $1.08$          & \cite{Delehaye2015}     & ENS Paris & 2015  \\
 ${}^{171}$Yb-${}^{87}$Rb   &  $0.16 \pm 0.02$             &  $2.4 \times 10^5$                      & $3.5 \times 10^5$   &     $2.00$               &  \cite{Vaidya2015}       &  University Maryland-NIST & 2015  \\
 ${}^{6}$Li-${}^{41}$K          &  $0.07$                            &  $1.5 \times 10^6$                      & $1.8 \times 10^5$   &      $2.23$            &  \cite{Yao2016}  & USTC Hefei and Shanghai & 2016  \\
\hline
\end{tabular}
\caption{Summary of some experimental efforts, in chronological order, towards achieving Fermi degenerate gases through sympathetic cooling. 
Reported is the adopted Fermi-Bose mixture, the minimum temperature achieved for the Fermi gas, the degeneracy factor $T/T_f$, the number 
of fermions at the deepest degeneracy $N_\mathrm{f}$, the quoted number of bosons near the end of the sympathetic cooling process $N_\mathrm{b}$, the Fermi-Bose 
trapping frequency ratio in the latest stage of sympathetic cooling, the related reference, the location of the laboratory, and the publication year of the cited paper. 
The acronym BDL indicates cases in which no discernible Bose cloud is observed or reported at the end of sympathetic cooling (Below Detection Limit).} 
\end{center}
\label{}
\end{table*}
The group reported $T_{\mathrm{Li}}/T_F=0.08 \pm 0.02$ ($T_{\mathrm{Li}}=290 \pm 3$ nK), while ${}^{174}$Yb at the same stage of cooling 
 has the temperature of $T_{\mathrm{Yb}}=280 \pm  20$ nK with the BEC critical temperature 
$T_c= 510$ nK, corresponding to $T_{\mathrm{Yb}}/T_c=0.55$. The number of atoms at the same stage was  
$N_{\mathrm{Li}}=2.5 \times 10^4$ and $N_{\mathrm{Yb}}=1.5 \times 10^4$. 

A group at University of Washington also reported achieving double degeneracy in the same mixture, with $T_{\mathrm{Yb}}/T_c=0.8$,  
$T_{\mathrm{Li}}/T_F=0.3$, $N_{\mathrm{Yb}}=2.3 \times 10^4$, and $N_{\mathrm{Li}}=1.2 \times 10^4$ \cite{Hansen2011}. 
This group optimized the spatial overlap between the clouds, a potential issue due to the large 
gravitational sagging, by using a magnetic field gradient acting only on ${}^6$Li \cite{Hansen2013}. 

\subsection{${}^{171}$Yb-${}^{87}$Rb}

This mixture has been studied at the University of D\"usseldorf in a hybrid trap consisting 
of an optical dipole trap for ytterbium and a magnetic trap for rubidium \cite{Tassy2010}, 
allowing independent trapping of the two species. Thermalization was studied for various ytterbium 
isotopes, and spatial separation for the Bose-Bose mixture  ${}^{174}$Yb-${}^{87}$Rb was also evidenced 
in the cold, but nondegenerate, regime \cite{Baumer2011}. 

More recently, a group at the Joint Quantum Institute, NIST, and the University of Maryland at College Park 
succeeded in achieving a degenerate Fermi-Bose mixture using a hybrid trap similar to the one of the D\"usseldorf group \cite{Vaidya2015}. 
Both these experiments are analyzed more extensively in the next section when discussing species-selective traps.

\subsection{${}^{40}$K-${}^{23}$Na}

 This mixture has been studied especially with the aim to exploit tunable interspecies interactions due to the 
expected presence of broad Feshbach resonances \cite{Park2012}, some occurring in low magnetic fields. 
The NaK molecule has a large permament dipole moment and is more stable against atom-atom exchange 
than, for instance, the KRb molecules. The apparatus is based on two independent Zeeman slowers,  
bringing both species to a common MOT. The atoms, after optical pumping into a stretched state of ${}^{23}$Na, 
are loaded into an optically plugged trap and then ${}^{23}$Na atoms are cooled with rf-induced evaporation. 
A degenerate Fermi gas of ${}^{40}$K with $2 \times 10^5$ atoms and $T/T_F=0.6$ is obtained, coexisting 
with a Bose-Einstein condensate of ${}^{23}$Na. 

The MIT group reported cooling trajectories with the phase space density versus the number of atoms of each species 
during various trapping stages, indicating a high evaporative cooling efficiency, attributed to the strong confinement in the 
optically plugged trap, and less efficient sympathetic cooling due to the large three-body collisions of ${}^40$K in the magnetic trap.
The lowest Fermi degeneracy factor $T/T_F=0.35$ is achieved with $3 \times 10^5$ ${}^{40}$K atoms after 
complete evaporation of the ${}^{23}$Na cloud. By using loss spectroscopy, more than 30 Feshbach resonances 
have been identified. 

\subsection{${}^{6}$Li-${}^{41}$K}

In a recent preprint, results on a ${}^6$Li-${}^{41}$K superfluid mixture have been reported \cite{Yao2016}. 
This case is complementary to the one explored in \cite{Ferrier2014,Delehaye2015}, because the mixture is 
made of species with a large mass imbalance, a situation of interest also for the physics of bound states of quarks 
\cite{Casalbuoni2004}. Instead of focusing on precision measurements of the center-of-mass motion as in 
\cite{Ferrier2014,Delehaye2015}, this experiment studied the formation of quantized vortices of the two species. 
Of particular relevance to our discussion is the adopted cooling technique, using ${}^{41}$K with sub-Doppler 
cooling in gray molasses, followed by two-stage evaporative cooling  in which ${}^6$Li is first 
sympathetically cooled by ${}^{41}$K in an optically plugged trap and then ${}^{41}$K is sympathetically cooled 
by ${}^6$Li in an optical dipole trap. Unlike other experiments described, the goal here is to have maximal 
quantum degeneracy with the maximum number of atoms for the two species and to have optimal contrast when 
imaging the quantized vortices of both species. This experiment has also succeeded in realizing a configuration 
with two-dimensional spin-orbit coupling, an important step toward exploring topological phases \cite{Huang2016}.

\subsection{Current experimental situation}

A partial list of experiments on quantum degenerate Fermi-Bose mixtures is given in Table I. 
It is evident that despite the diversity and ingenuity of experimental techniques, the 
progress in reaching lower $T/T_F$ combined with a large number of fermions is rather slow. 
This can jeopardize the chances of achieving an interesting regime in which model Hamiltonians can 
be simulated, unless the reasons for these limitations are understood and mitigated or circumvented. 
The main goal of the next section is to identify basic limitations to fermion cooling and to present 
a set of techniques, either implemented or mature enough to be implemented, potentially enabling 
the deepest Fermi degeneracies.

\section{Experimental techniques to achieve deeper Fermi degeneracy} 

We report here a subset of proposals to reach a deeper quantum degeneracy, with major emphasis on 
their practical feasibility. Specifically, we focus attention on trapping setups with a stage of 
sympathetic cooling in which the two species can be trapped with different confinement strengths, frictionless 
cooling techniques, and use of reduced dimensionality. A separate class of cooling techniques is also currently under 
development for single species, based on all-optical trapping and cooling, as described in Section VI D.  

\begin{figure}[t]
\includegraphics[width=0.95\columnwidth]{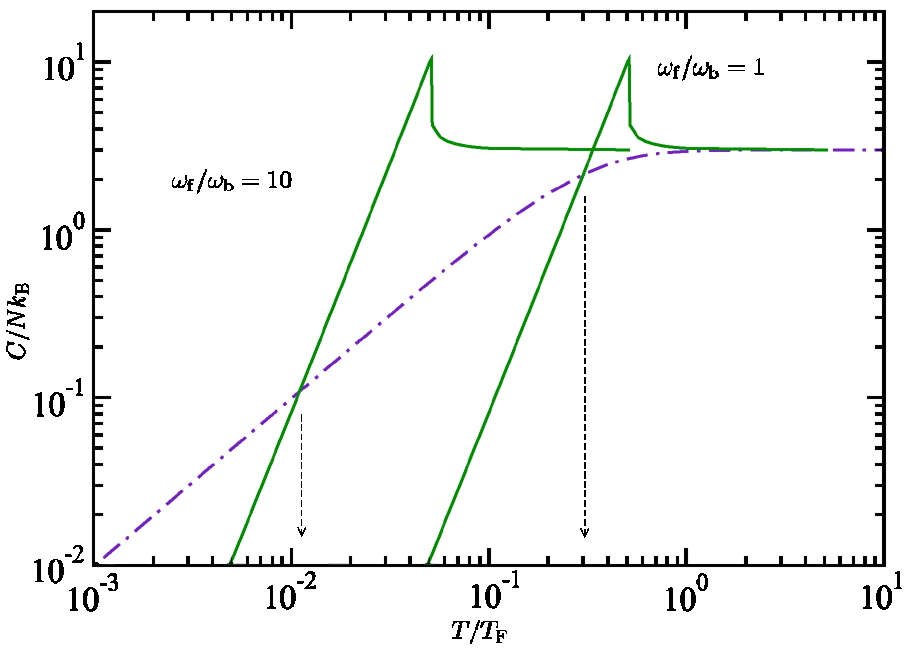}
\caption{Heat capacity mismatching in Bose-Fermi mixtures. The specific heat of noninteracting fermions (dashed-dotted line) and bosons 
(solid line) is shown versus temperature for two different values of the trap frequency ratio $\omega_\mathrm{f}/\omega_\mathrm{b}$.
The crossover between the two curves occurs at $T/T_F$ values indicated by arrows, and smaller values of $T/T_F$ correspond to 
progressively more inefficient sympathetic cooling of the Fermi gas. It is evident that a tenfold increase in the trapping frequency ratio 
between the fermionic and the bosonic species results in a tenfold decrease in the $T/T_F$ at which the heat capacity crossover occurs. 
The specific heat is calculated numerically from first principles for a finite number of fermions and bosons, both equal to  $10^6$, of equal 
mass as approximately occurring in isotopic mixtures,$m_\mathrm{b}=m_\mathrm{f}$, and trapping frequencies 
$\omega_x=\omega_y=\omega_z/\sqrt{2}$ as achievable in a crossed dipole trap (reproduced from \cite{Presilla2003}).}
\end{figure}

\begin{figure*}[t]
\begin{center}
\includegraphics[width=0.32\textwidth]{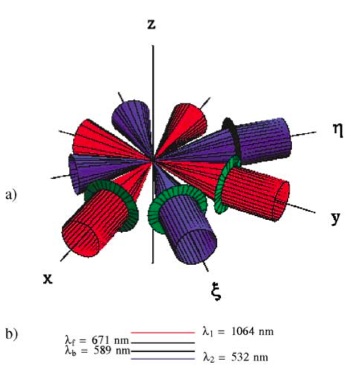}
\includegraphics[width=0.32\textwidth]{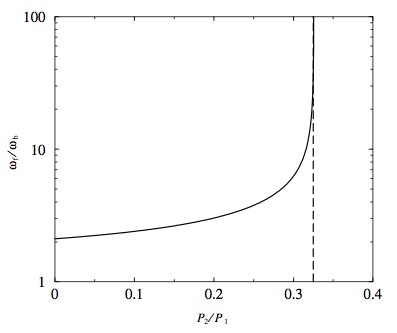}
\includegraphics[width=0.32\textwidth]{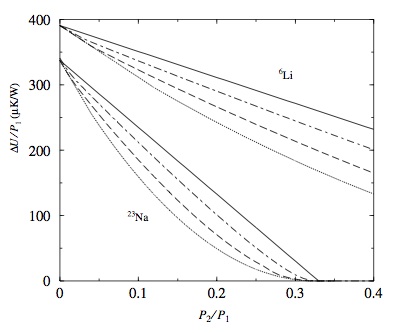}
\caption{(left) (a) Geometry of species-dependent optical dipole traps using selectively deconfining laser beams, in the example of an
${}^{6}$Li-${}^{23}$Na Fermi-Bose mixture and with the use of a laser emitting at the wavelength of 1064 nm and its second harmonic at 532 nm. 
(b) Schematics of the involved atomic and laser wavelengths. 
(center) Trapping angular frequencies ratio for the fermionic and bosonic species versus  the power ratio of deconfining and confining laser beams. 
We note that $\omega_\mathrm{f}/\omega_\mathrm{b}>1$ for $P_2/P_1=0$ as a consequence of the mass ratio between $^6$Li and $^{23}$Na. 
We assume equal waists for the Gaussian beams, and  the spontaneous emission linewidths $\Gamma_\mathrm{f}=2 \pi \times 5.9~\mathrm{MHz}$, 
$\Gamma_\mathrm{b}=2 \pi \times 9.8~\mathrm{MHz}$,  for $^6$Li and $^{23}$Na, respectively. 
The dashed line indicates the critical value of the laser beams power ratio $P_2/P_1$ for which the bosons are deconfined. For values of 
$P_2/P_1$ smaller and in the neighborhood of this critical value, the bosons are so weakly confined that relative gravity sagging between the 
Fermi and Bose species can suppress sympathetic cooling due to the marginal spatial overlap. (right) Confining energy $\Delta U$ per unit of infrared laser power for the 
fermionic and bosonic species as a function of the laser beam power ratio. For each species, from top down, we show the curves obtained with the blue-detuned beams 
rotated with respect to the red-detuned ones by $\theta=0$ (coaxial case), $\pi/16$, $\pi/8$, and $\pi/4$, respectively (reproduced from \cite{Onofrio2002}).}
\end{center}
\end{figure*}

\subsection{Species-selective trapping}

The mismatching of the heat capacities during sympathetic cooling of a Bose-Fermi mixture has been perceived from the very 
beginning as the major obstruction to achieving the lowest values of the Fermi degeneracy parameter, as first discussed 
in \cite{Schreck2001a,Truscott2001}.  From the theoretical standpoint, the problem corresponds to solving coupled 
Boltzmann equations for the two species in the presence of a confining potential, interatomic and intratomic interactions, 
including quantum statistics in the degenerate regime \cite{Geist1999}. 
In \cite{Viverit2001}, it was proposed to bring a Fermi gas to deeper degeneracy by superimposing a deep and narrow 
optical dipole trapping potential to on the broader magnetic trapping potential. 
The stiffer potential allows reaching a deeper Fermi temperature in the proximity of the trap minimum, analogously 
to the successful demonstration of a reversible BEC  \cite{Stamper1998}. 
However, this enhancement of the Fermi temperature involves only a small fraction of the Fermi gas, which 
is detrimental to the achievement of a signal-to-noise ratio large enough to observe the relevant physics. 
In \cite{Wouters2002} the idea, whose implementation was already begun in the ENS and Rice experiments, of also evaporating  
fermions was developed more quantitatively by modeling the dynamics of alternating stages of 
sympathetic cooling in a Bose-Fermi mixture and evaporative cooling of fermions. 
Evaporating fermions obviously decreases their number, similarly ending up with limitations on the signal-to-noise 
ratio for detecting interesting physics, and also decreases the corresponding Fermi temperature. 

Other studies focused on the limits of sympathetic cooling in the presence of particle losses, for instance 
as discussed in \cite{Timmermans2001}, leading to Fermi degeneracy factors related to the ratio between the 
loss rate and the elastic collision rate of the Fermi and the Bose species \cite{Idziasek2005}. 

An alternative cooling model, not representative of usual sympathetic cooling proceeding via quasi-equilibrium states between the 
Fermi and the Bose gases, was discussed in \cite{Carr2004a}. In this model, the Bose gas is driven to 
the lowest temperature using evaporative cooling, without interactions with the Fermi gas left at higher temperature (for instance, by 
tuning their interspecies Feshbach resonance to achieve a zero elastic scattering length), and then the two gases are suddenly 
interacting. The excited bosons are removed and the Bose gas is cooled again. In a subsequent paper \cite{Carr2004b}, the role of the 
heat capacity of the Bose gas was discussed using a definition of cooling efficiency based on the temperature drop of the 
Fermi gas per unit of evaporated bosons.

A solution overcoming the drawbacks present in the former proposals has been put forward in \cite{Onofrio2002}. 
The heat capacity matching is hindered by the decreasing heat capacity of the Bose gas when it enters the degenerate 
regime because the Bose-condensed phase has no heat capacity, and the thermal component has a heat capacity independent, 
in the non-interacting case, of the number of thermal bosons. Therefore, one can mitigate this issue by forcing the 
Bose gas to enter the quantum degenerate regime as late as possible with respect to the Fermi gas, to preserve its cooling capability to the 
maximum extent. Since the critical temperature for Bose-Einstein condensation and the Fermi temperature 
are directly proportional to the trapping frequencies, this can be achieved by using a large trapping ratio $\omega_f/\omega_b$. 
An alternative standpoint corroborates this view, as this is equivalent to maintaining the Bose gas as classical as possible, which 
implies making $\hbar \omega_b$ as small as possible compared with the corresponding energy quantum $\hbar \omega_f$ of the Fermi gas. 

An easy way to increase the $\omega_f/\omega_b$ ratio is available by conveniently choosing atomic masses and hyperfine states 
in a magnetic trap \cite{Brown2004}. In particular, it seems advantageous to choose the lightest fermion and the heaviest boson, and 
in the specific case of magnetic traps to choose hyperfine states for which the Land\'e g-factors of the fermion and the boson are 
respectively maximum and minimum. However, this does not allow much flexibility and makes the mixture prone to other limitations 
like relative gravitational sagging, resulting in an intermediate value of the mass ratio optimizing sympathetic cooling. 

An alternative approach is to use a bichromatic optical dipole trap. The original motivation discussed in \cite{Onofrio2002} was 
related to the possibility of increasing the Fermi velocity compared with the critical velocity for the breakdown of superfluidity in the 
Bose gas, avoiding the suppression of scattering expected in a Bose-Fermi mixture \cite{Timmermans1998}. 
The main gain in achieving species-selective trapping -- a better heat capacity matching between the 
two species -- was later discussed in detail in \cite{Presilla2003}.
A strategy to improve cooling is achieved by shifting the crossover point of the heat capacities to the lowest $T/T_F$ as much as possible, and 
this may be achieved by increasing the ratio $\omega_\mathrm{f}/\omega_\mathrm{b}$. 
For instance, in Fig. 4 we show that for $\omega_\mathrm{f}/\omega_\mathrm{b}=10$, the heat capacity inversion 
takes place at $T/T_\mathrm{F}\simeq 10^{-2}$, instead of $T/T_\mathrm{F}\simeq 0.3$ in the case of equal trapping frequencies. 
Different trapping potentials for the two species in an optical dipole trap can be engineered by using proper detuning and intensities of two laser beams. 
In a usual optical dipole trap, the laser wavelength is red-detuned with respect to the atomic 
transitions, resulting in an effective attractive potential for both the atomic species. 
If a second beam with the wavelength blue-detuned only with respect to the Bose species is added, the 
confinement of the latter is weakened with respect to the one for the Fermi species. 
The stronger confinement for the fermionic species implies that the degeneracy 
condition for it is met earlier than for the more weakly confined Bose species. 

In the specific example of a ${}^6$Li-${}^{23}$Na mixture considered in \cite{Onofrio2002}, a 
crossed-beam geometry is adopted for two different wavelengths, and the two laser wavelengths 
are chosen at $\lambda_1=1064$ nm and $\lambda_2=532$ nm, for instance by using a Nd:YAG laser 
and a frequency-doubling crystal. The relevant atomic transition is at 
$\lambda_\mathrm{b}=589$ nm for sodium and at $\lambda_\mathrm{f}=671$ nm for lithium. 
Compared with sodium atoms, the lithium atoms are closer to the (attractive) red-detuned laser at 1064 nm 
and farther from the (repulsive) blue-detuned laser at 532 nm.     

With reference to the set up depicted in Fig. 5, the effective potential energy felt by an atom of a species $\alpha$ ($\alpha=\mathrm{b}$ 
for $^{23}$Na and $\alpha=\mathrm{f}$ for $^{6}$Li) and due to a laser beams $i$ ($i=1,2$) is \cite{Ashkin1979}
\begin{equation}
U_{i}^{\alpha}(x,y,z)=-\frac{\hbar \Gamma_{\alpha}^2}{8 I_{\alpha}^\mathrm{sat}}
\left( \frac{1}{\Omega_\alpha - \Omega_i} +\frac{1}{\Omega_\alpha + \Omega_i} \right)  I_i(x,y,z),
\end{equation}
where $\Gamma_{\alpha}$ is the atomic transition linewidth, $\Omega_\alpha=2 \pi c/\lambda_{\alpha}$, $\Omega_i=2 \pi c/\lambda_i$, 
$I_i$ is the laser intensity, and $I_{\alpha}^\mathrm{sat}$ is the saturation intensity for the atomic transition, expressed in terms of the former 
quantities as $I_{\alpha}^\mathrm{sat}=\hbar \Omega_\alpha^3 \Gamma_\alpha/(12 \pi c^2)$.
Each laser intensity $I_i$ is the incoherent sum (obtained by proper polarization or a relative detuning of the orthogonal beams) of the 
intensities of the two beams propagating along orthogonal directions in the $xy$ plane and focused at $(x,y,z)=(0,0,0)$.
According to Fig. 5a, we assume that the red-detuned beams propagate along the axes $x$ and $y$, while the blue-detuned 
ones along the axes $\xi$ and $\eta$ rotated with respect to $x$ and $y$ by an angle $\theta$, $\xi=x\cos\theta + y\sin\theta$ 
and $\eta=y\cos\theta - x\sin\theta$, with $0\leq \theta \leq \pi/4$. In both cases, we can write 
\begin{eqnarray}
I_i(x,y,z) &=& 
\frac{2 P_i}{\pi w_i^2 \left(1+{\xi^2}/{R_i^2}\right)}
\exp\left[
-\frac{2(\eta^2+z^2)}{w_i^2 \left(1+{\xi^2}/{R_i^2}\right)}\right]
\nonumber \\
&+&
\frac{2 P_i}{\pi w_i^2 \left(1+{\eta^2}/{R_i^2}\right)}
\exp\left[
-\frac{2(\xi^2+z^2)}{w_i^2 \left(1+{\eta^2}/{R_i^2}\right)}\right],~~~
\end{eqnarray}
where $P_i$ is the beam power, $w_i$ is the $1/e^2$ beam waist, and $R_i=\pi w_i^2/\lambda_i$ is the Rayleigh range.
The total potential experienced by the fermions (bosons) is $U_\mathrm{f}=U_1^\mathrm{f}+U_2^\mathrm{f}$ 
($U_\mathrm{b}=U_1^\mathrm{b}+U_2^\mathrm{b}$). When $P_1/P_2$ is large enough, both potentials 
$U_\mathrm{f}$ and $U_\mathrm{b}$ have a minimum at $(x,y,z)=(0,0,0)$, and both $T_F$ and $T_c$ are 
determined by the small oscillation frequencies around this minimum.
Neglecting the terms $[\lambda_i/(\pi w_i)^2]$ compared with unity, we find
\begin{eqnarray}
\omega_{\alpha x} = \omega_{\alpha y} = \frac{\omega_{\alpha z}}{\sqrt{2}} = \sqrt{\frac{\hbar}{\pi m_\alpha}
\left(\frac{k_1^\alpha P_1}{w_1^4}+\frac{k_2^\alpha P_2}{w_2^4}\right)},
\label{omega}
\end{eqnarray}
where $m_\alpha$ is the mass of an atom of the species $\alpha$ and
\begin{eqnarray}
k_{i}^{\alpha} = \frac{\Gamma_{\alpha}^2}{I_{\alpha}^\mathrm{sat}}
\left( \frac{1}{\Omega_\alpha - \Omega_i} +\frac{1}{\Omega_\alpha + \Omega_i} \right).
\end{eqnarray}

The trapping angular frequencies in Eq. (\ref{omega}) are independent of the angle $\theta$ between the 
blue- and red-detuned laser beams. This strategy can also be applied to other mixtures such as 
$^6$Li--$^{87}$Rb, for which $\lambda_\mathrm{b} > \lambda_\mathrm{f}$, by choosing a blue-detuned beam 
wavelength such that $\lambda_\mathrm{f} < \lambda_2 < \lambda_\mathrm{b}$, also including the use of far-off 
resonant trapping through a 10.6 $\mu$m CO${}_2$ laser \cite{Onofrio2004}.

\begin{figure}[t] 
\begin{center}
\includegraphics[width=0.95\columnwidth]{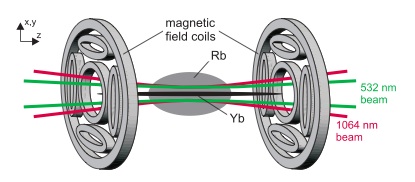}
\caption{Schematics of the species-selective trapping for the ${}^{87}$Rb-${}^{133}$Yb 
mixture of the group at the University of D\"usseldorf, with the independent trapping of 
rubidium in a magnetic trap and ytterbium in a bichromatic optical dipole trap designed 
to cancel the light shifts induced on rubidium (reproduced from \cite{Baumer2011}).}
\end{center}
\end{figure}

\begin{figure}[t]
\begin{center}
\includegraphics[width=0.95\columnwidth]{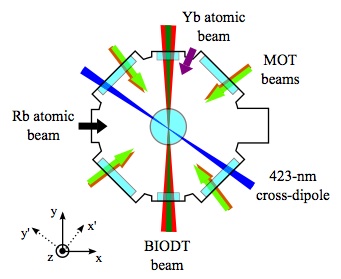}
\includegraphics[width=0.99\columnwidth]{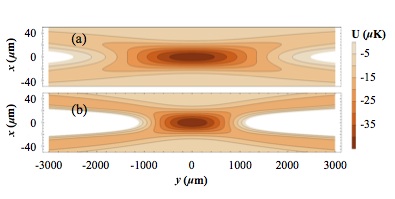}
\caption{(Top) Schematics of the apparatus for species-dependent trapping of ${}^{87}$Rb and 
${}^{171}$Yb. A bichromatic optical dipole trap (BIODT) with confining and deconfining laser 
beams operating at respective wavelengths of 1064 nm and 532 nm is used to confine ${}^{171}$Yb 
and to cancel the optical dipole trapping potential for ${}^{87}$Rb, which is independently confined 
in a magnetic trap. Further axial confinement of ${}^{171}$Yb is ensured by a 423 nm laser beam 
in a crossed-beam configuration with the BIODT, forming an angle of 57${}^0$. The atoms are loaded 
via two Zeeman slowers operating independently, and precooled in a combined MOT.  
The vertical direction is orthogonal to the laser beams, along the $z$ axis.
(Bottom) Contour curves for the expected potential energy experienced by ${}^{87}$Rb atoms in the 
BIODT trap for two different values of the power in confining and deconfining laser beams. 
Case (b) is for $P_{1064}$=0.8 W, $P_{532}$=1.4 W, corresponding to $P_2/P_1=1.75$ in the notation 
used in Fig. 5, while case (c) is for $P_{1064}$=2.1 W, $P_{532}$=5.0 W corresponding to $P_2/P_1=2.38$. 
Accordingly, in the second case, the trapping region shrinks in side and the overall potential depth 
(reproduced from \cite{Vaidya2015}).}
\end{center}
\end{figure}

Species-selective trapping based on bichromatic optical dipole traps has been the subject of extensive experimental 
investigations by two groups, in D\"usseldorf \cite{Tassy2010,Baumer2011} for a ${}^{87}$Rb-${}^{174}$Yb Bose-Bose mixture, 
and more recently at NIST in Gaithersburg \cite{Vaidya2015} for a ${}^{87}$Rb-${}^{171}$Yb Fermi-Bose mixture. 
In both cases a bichromatic trap created by superimposing single laser beams at the wavelengths of 532 nm 
and 1064 nm is used to trap ytterbium, for which both beams create a confining potential (see Figs. 6 and 7).
The laser powers at the two wavelengths are adjusted in such a way that the confining (for 1064 nm) 
and deconfining (for 532 nm) potentials are cancelled for ${}^{87}$Rb at least in the trap minimum and its vicinity, 
where the harmonic approximation of the Gaussian potentials is valid. This allows independent 
trapping of the two species, because ${}^{87}$Rb is confined by means of a magnetic trap. 
It is worth noting here that the NIST group has achieved the degeneracy factor  $T/T_F=0.16 \pm 0.02$, with trapping frequencies along the three axes, 
at the latest stage of cooling, with $(\omega_x,\omega_y,\omega_z)=(150,140,75)$ Hz for ${}^{171}$Yb 
and $(\omega_x,\omega_y,\omega_z)=(140,140,10)$ for ${}^{87}$Rb, corresponding to the ratio
$\omega_f/\omega_b=2$, quite favorable from the standpoint of heat capacity matching.
In principle, the trapping frequency ratio between the two species can be changed, and the dynamics of evaporative and 
sympathetic cooling studied along the lines suggested in \cite{Presilla2003,Onofrio2004,Brown2008}. 
The added flexibility in independently controlling the trapping potentials for the two species comes at a price, however, because  
sympathetic cooling is quite sensitive to the spatial overlap. In \cite{Vaidya2015}, a procedure for the optimization of the 
overlap between the two clouds based upon displacing the magnetic trap and monitoring sympathetic heating 
of the ytterbium cloud was also discussed. 

\begin{figure*}[t]
\vspace{-4.5cm}
\begin{center}
\includegraphics[width=0.95\textwidth]{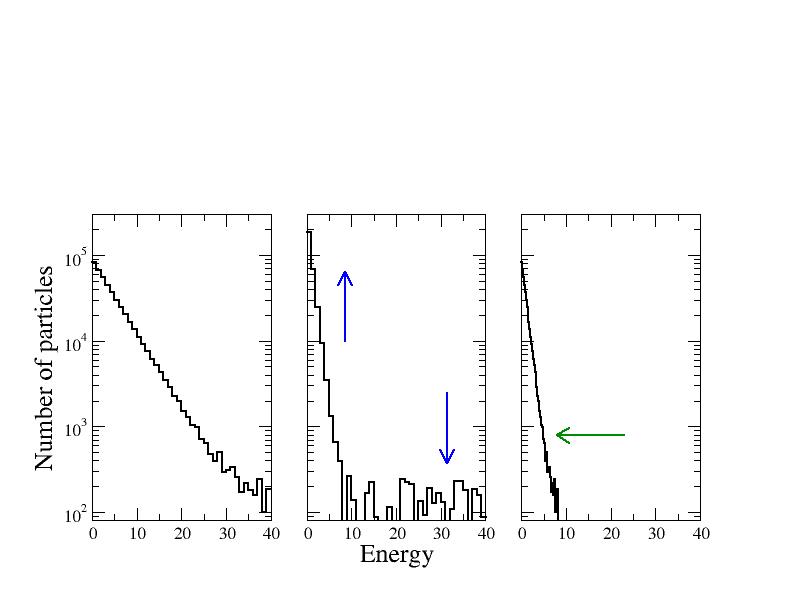}
\caption{Schematics of two possible cooling schemes, in the simplest case of a system of one-dimensional quantum harmonic oscillators 
with an angular frequency $\omega$  and the Boltzmann energy distribution.  The left plot shows the energy distribution, in the form of a histogram 
in which the number of occupied states at energies between $E_k$ and $E_k+\Delta E$ (in logarithmic scale, with $\Delta E=\hbar \omega$) 
is reported versus $E$, at a generic initial temperature $T_i$. The goal is to bring the state to a new Boltzmann distribution pertaining to 
another temperature $T_f< T_i$. Shown in the middle plot is the usual way to achieve this goal, by reshuffling energy such that 
higher-energy states become depopulated by transferring particles to lower-energy states, as indicated by the blue arrows. 
In the right plot, the case of adiabatic cooling is shown instead, for which the populations do not change, while the energy scale decreases by 
a constant factor, resulting in a global shift of the energy distribution to the left, as indicated by the green arrow. 
The final outcome in terms of final temperatures is the same, as indicated by the identical slope of the two distributions, and in both cases energy is  
released to the external environment.  For the population-changing case in the middle plot, the energy release must occur because the more energetic particles 
need to give up some energy to be demoted to the lower energy states, in the right plot because an external agent has decreased the overall 
energy scale of the system by a constant scaling factor. Unlike the first cooling technique, the multiplicity of each energy level in the second one is unchanged, 
as is the Shannon entropy of the system; hence the name adiabatic cooling. Analogous considerations can be discussed in the case 
of heating processes following the same protocols, but reversed in time. In this example, we assume a system with $4\times 10^5$ oscillators, the final 
temperature $T_f=T_i/5$, and $\Delta E=T_i/40$.}
\label{}
\end{center}
\end{figure*}

\subsection{Frictionless adiabatic expansion or compression}

An alternative cooling method is available by exploiting adiabatic expansion of the atomic species, a technique 
rather simple to  explain qualitatively. In Fig. 8, we show an initial Boltzmann distribution (left plot) and two ways to reduce the associated 
temperature, in the middle and right plots. In the middle plot, population transfer occurs from higher-energy states to lower-energy states. 
This is achieved by subtracting energy and entropy in the system; even graphically, it is easy to check that both are decreased in the final 
configuration, the former due to the progressing depletion of high-energy states and the latter because the multiplicity of states decreases  
as particles bunch in fewer energy levels. It is possible to alternatively envisage a cooling scheme in which the particle population is unchanged, and 
instead the energy axis is simply scaled down, as in the right plot of Fig. 8. The overall energy is then changed by an amount equal to the scaling factor, 
while the entropy is left unchanged, and we hence use the term adiabatic. Usually, this dynamics requires long timescales, because otherwise 
excitations would occur, upsetting the particle population. Nevertheless, there are cases where a carefully chosen time dependence of the scaling factor 
allows an overall adiabatic process, even if the atomic population is changed at intermediate times. Adiabatic cooling has been used to achieve 
the lowest temperature measured in a laboratory by using a Bose condensate \cite{Leanhardt2003}. However, the lower density achieved after 
the expansion compensates the temperature drop, eventually yielding the same phase space density, implying no net gain in quantum degeneracy \cite{Ketterle1992}. 

\begin{figure*}[t]
\begin{center}
\includegraphics[width=0.49\textwidth]{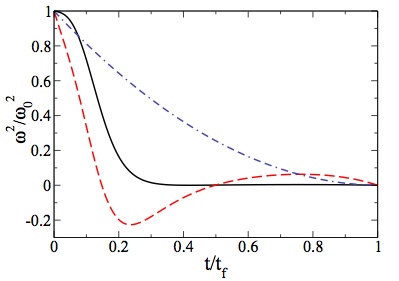}
\includegraphics[width=0.46\textwidth]{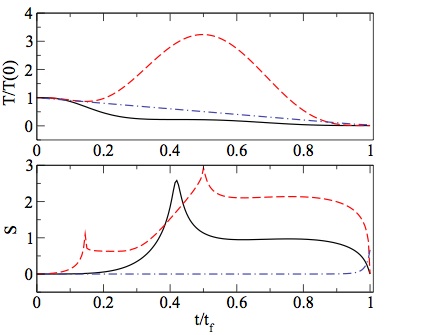}
\end{center}
\caption{Example of protocols for fast frictionless cooling of an atomic species and corresponding 
evolution of thermodynamic quantities.  (Left)  Various decompression strategies 
are characterized by the square of the trapping frequency, normalized to its initial 
value, versus time normalized to the overall duration of the decompression. 
The curves here indicate an Ermakov trajectory with positive square frequency lasting 
25 ms (black continuous line), a fast Ermakov trajectory of total duration of 6 ms 
(red dashed line), and a linear ramp-down from the trapping frequency for the usual slow 
adiabatic strategy (blue dashed-dotted line) with a duration of 400 ms. 
For the intermediate strategy with the 6 ms duration, the squared frequency becomes negative in a finite interval.
(Right, top) Temperature scaled to initial temperature versus time. For the fastest 
decompression strategy lasting 6 ms, the temperature exceeds the initial temperature in 
a time interval related to the one in which the squared frequency becomes negative, unlike 
the case of the strategy lasting  25 ms, in which the temperature decreases monotonically. 
The presence of an intermediate temperature exceeding the initial one occurs for all the protocols 
in which the time duration is shorter than 11 ms. (Right, bottom) Time dependence of the Shannon 
entropy, defined as $S(t) = - \sum_{n}  |c_{n}(t)|^2 \log |c_{n}(t)|^2$, with $c_{n}(t)$ the 
coefficients of the wave function in the instantaneous eigenstate basis.
The final entropy exactly equals the initial one in the frictionless cooling cases, even if 
it changes in time at intermediate stages. Conversely, the linear decompression 
has a nearly constant entropy throughout the evolution except at the end, when the process 
can no longer be considered adiabatic for any ramp-down of the frequency (reproduced from \cite{Choi2011}).}
\end{figure*}

In the framework of atomic mixtures, we can envisage a situation where one species (the 'target') is compressed while its temperature 
is unchanged, and the other species (the 'bath') has a much larger heat capacity. In a complementary and equivalent approach, we can 
imagine keeping the target species unaffected, but instead expanding the bath species with a consequent decrease in the common temperature. 
Even in this second situation, the net effect is an increase in phase space, this time due to the lower common final temperature rather than  
due to the higher density of the target species as in the former case. Due to the need to maximize the spatial overlap between the two species for 
efficient thermalization, this calls for confining the two species in the same region of space with different, and independently controllable, trapping strengths. 
Full optimization of this cooling strategy is achieved if the bath species always retains the maximum value of its heat capacity. 
For a Bose gas acting as a bath, this is achieved if its initial state corresponds to a nondegenerate regime or, although it is difficult to achieve, 
if it always kept at its critical temperature for the BEC, exploiting the peak of the specific heat. 

\begin{figure*}[t]
\begin{center}
\includegraphics[width=0.95\textwidth]{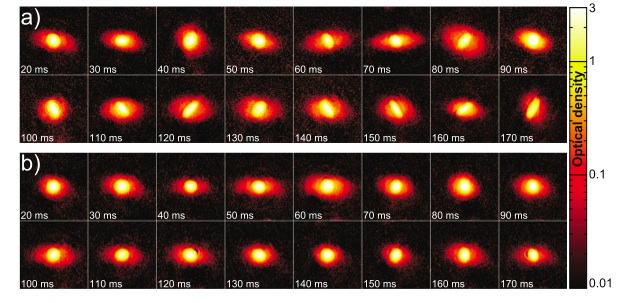}
\caption{Experimental comparison of two decompression strategies, one based on a linear ramp-down of the trapping 
frequency (top image set, a), the other based on a shortcut to the adiabatic trajectory (bottom image set, b), both with 
a duration of 30 ms, for a ${}^{87}$Rb BEC. The times indicated in each image are the hold times 
in the trap after the decompression strategy and before the release for the absorption imaging picture taken after a time 
of flight of 28 ms. The field of view of each image is 545 $\mu$m $\times$ 545 $\mu$m. The dipole oscillation of the 
center of mass has been subtracted in each picture to emphasize the dynamics of the quadrupole motion, and its 
amplitude is also reduced as in the visible case of the quadrupole oscillation (reproduced from \cite{Schaff2011NJP}).}
\end{center}
\end{figure*}

The usual adiabatic expansion of a gas can be limited by the speed of the process, but the use of a technique called shortcut to adiabaticity 
or fast frictionless cooling to achieve the same expansion on a much shorter timescale has been introduced in atomic physics \cite{Chen2010,Torrontegui2013}. 
This has led to implementations for fast decompression of ${}^{87}$Rb atoms in both nondegenerate \cite{Schaff2010} and degenerate 
\cite{Schaff2011,Schaff2011NJP} regimes, and to proposals for fast atomic transport \cite{Torrontegui2011}.
The case of the harmonic oscillator is particularly simple due to the equidistance of the energy levels, which can then  be changed by a common 
factor by simply changing the angular frequency. If this is done quickly, excitations are generated; however, due to the commensurability 
among all the energy eigenvalues, there are times for which the initial population is revived, with the final entropy equal to the initial one. 
The usual Hamiltonian operator is superseded by the Lewis-Riesenfeld operator defined as \cite{Lewis1967,Lewis1969} 
\begin{equation}
\hat{I}(t)=\frac{\hat{\pi}^2}{2m}+ \frac{m \omega_0^2 \hat{q}^2}{2 b^2},
\end{equation} 
where $\hat{\pi}=b\hat{p}-m \dot{b}\hat{q}$ is the momentum operator conjugate to the operator $\hat{q}/b$. 
The parameter $\omega_0$ can be chosen as the initial frequency, and $b(t)$ is a time-dependent frequency scaling factor, which, in 
order that $\hat{I}(t)$ be an invariant operator, must satisfy the Ermakov equation \cite{Ermakov1880} $\ddot{b}(t) + \omega(t)^2b(t) = \omega_0^2/b^3(t)$,  which 
can be solved by imposing boundary conditions on $b(t)$ and its first and second time derivatives, and properly choosing $\omega(t)$.  
For a targeted final trapping frequency $\omega_{\mathrm f}$ and a time duration $t_{\mathrm f}$ for the process \cite{Chen2010},  an Ermakov trajectory 
is obtained if the trapping frequency satisfies the equation \cite{Chen2010,Torrontegui2013}
\begin{equation}
\omega^2(t) = \frac{\omega_0^2}{b^4(t)} -  \frac{\ddot{b}(t)}{b(t)}.
\end{equation} 

Of particular relevance is the possible presence of an antitrapping stage corresponding to $\omega^2(t)<0$ at intermediate 
times if the time duration for the process is shorter than a threshold value \cite{Chen2010}. This enables the wave function to 
spread out faster so as to reach the final target width within the targeted time duration. In a realistic application, this  
also implies temporary heating of the atomic clouds, which could end in their loss whenever the realistic 
confinement potential has a finite depth. Experimentally, an antitrapping stage can be obtained by properly time-modulating 
the power of  blue-detuned beams in an optical trap configuration similar to the one discussed in Section VI. A. 

It has been proposed to use fast frictionless cooling with a Fermi-Bose mixture in which the Bose 
gas is adiabatically expanded starting from a nondegenerate, classical state \cite{Choi2011}. 
Compared with the usual evaporative cooling of a Bose gas, a further advantage is that the 
number of bosons is constant, keeping the heat capacity constant during the process. 
Figure 9b shows the maximum achieved temperature, scaled to the initial temperature and 
expressed in units of quanta of the corresponding instantaneous harmonic oscillator 
$k_B T_\mathrm{max}/\hbar \omega(t_\mathrm{max})$, versus the time duration of the 
cooling procedure.  This figure of merit is useful in  assessing trap losses in the case 
of a finite trap depth. The results show that for $t_{\mathrm f}$ shorter than approximately 
$11$ ms, the maximum temperature of the atoms becomes comparable to the energy of the 
15${}^\mathrm{th}$ excited state of the instantaneous harmonic trap, and rapidly increases 
for shorter $t_{\mathrm f}$. Nevertheless, it appears that efficient cooling  with the minimal 
trap loss on timescales comparable to the oscillation period of the trap is possible as 
long as excessively short cooling times are avoided. This is clear from the temperature curve, 
where for $t_{\mathrm f} > 11$~ms the initial temperature is also the highest temperature, indicating
stability in the dynamics as the trap is made more shallow. By contrast, at shorter times,
substantial heating resulting from the antitrapping region would make fast frictionless
strategies difficult to implement. The maximum heating due to antitrapping was found to 
occur typically at  $t = t_{\mathrm f}/2$,  the time when the trap changes its curvature 
at the end of antitrapping. A sudden jump occurs at around  $t_{\mathrm f}  \approx 11$ ms, 
because in this case the maximum heating due to antitrapping has the same magnitude as the 
initial temperature.  Because antitrapping occurs for all $t_{\mathrm f}< 25$ ms in this example, the 
presence of an antitrapping stage alone is not a sufficient condition for heating the atoms to 
temperatures greater than the initial temperature.

\begin{figure}[t]
\begin{center}
\includegraphics[width=0.95\columnwidth]{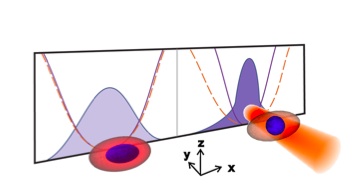}
\includegraphics[width=0.95\columnwidth]{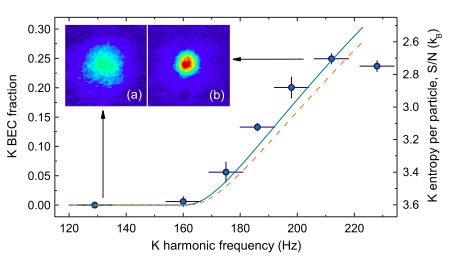}
\caption{Schematic of the species-selective trapping for the ${}^{39}$K-${}^{87}$Rb 
mixture of the Florence group (top) and dynamics of compression (bottom) in which the condensate 
fraction of ${}^{40}$K is shown to progressively increase with the increase in the trapping frequency 
owing to the increase of the laser power in the optical dipole trap (reproduced from \cite{Catani2009}).}
\label{}
\end{center}
\end{figure}

A different issue arises when the spatial overlap of the two clouds is taken into account, 
as the coolant species considerably increases its size during its decompression. 
As discussed in \cite{Chen2010}, the position variance is directly related to the 
frequency scaling factor $b$ as $\sigma_{x}^2 = \hbar (n+1/2) b^2/(m \omega_0)$.
For the two adiabatic invariant strategies,  the temporal variation of the position variance is 
independent of $t_{\mathrm f}$, as expected from the fact that  the position variance is proportional 
to $b(t)$. Because $\sigma^{2}_x(t_{\mathrm f})/\sigma^{2}_x(0) = b^{2}(t_{\mathrm f})/b^{2}(0) = 
\omega_0/\omega_{\mathrm f} = 10^2$ as a result of the boundary condition for $b(t)$, it is 
conceivable that, if a large $b(t_{\mathrm f})$ is targeted, the spread of the atomic cloud 
of the coolant would result in a small overlap with the cloud to be sympathetically cooled.  
The linear ramp-down result also exhibits a large broadening expected of an adiabatic process 
that relaxes the trap frequency by a factor of $10^2$. The overlap issue seems to be crucial 
in various experiments (see, e.g., \cite{Hansen2011} for the effect of gravitational sagging 
on a large mass ratio Fermi-Bose mixture), and its effect on species-selective traps has been 
discussed in detail in \cite{Brown2008}. 

We note that in practice, the issue of spatial overlap is less severe than expected, since by pure chance the 
fermionic isotope of the alkali primarily used in the experiments, ${}^{6}$Li, is lighter than the bosonic species used as 
coolants, ${}^{23}$Na, ${}^{87}$Rb, and bosonic isotopes of Yb. Therefore, the initial overlap 
between the two species sees the fermionic species more spread out than the bosonic species. 
The decompression of the bosonic species alone in general allows a better spatial overlap 
at intermediate times, but this also provides a limit on the maximum allowed decompression 
before the spatial overlap decreases again. It is also worth pointing out that, especially 
for very short cooling times, a possible issue arises with the sympathetic equilibration rate 
between the two species, depending on the interspecies elastic scattering rate. This seems 
less relevant than the spatial overlap for most of the concrete implementations, and can 
be circumvented by using magnetic or optical Feshbach resonances to boost the elastic 
scattering length.

On the experimental side, this technique seems mature to be implemented for fermion cooling, but 
so far only disjoint elements have been demonstrated. A group at the University of Nice-Sophia Antipolis 
has studied the dynamics of fast decompression of an ultracold cloud of ${}^{87}$Rb in a magnetic trap. 
A final trap frequency 15 times smaller than the initial trap frequency was achieved in 35 ms, with a strong 
suppression of sloshing and breathing modes induced during the transient by both the vertical displacement and the 
decrease in curvature with respect to a standard compression consisting in a linear ramp-down 
of the trapping frequency in a time 37 times longer \cite{Schaff2010}. 
The experiments were repeated for an interacting BEC of ${}^{87}$Rb (Fig. 10), obtaining 
similar results for both the condensate and the thermal fraction, showing the universality 
of the chosen trapping frequency trajectory \cite{Schaff2011,Schaff2011NJP}. 
The validity of the Ermakov construction for designing shortcuts to adiabaticity for BECs 
is related to the fact that the time evolution of the macroscopic wavefunction in a time-dependent harmonic 
potential is a dilatation, a scaling property first discussed in \cite{Castin1996}.

These demonstrations of shortcuts to adiabaticity were implemented for a single species of a Bose gas.
In the case of a two-species setup, it can be decided to increase the phase-space density of the 
target species by expanding the coolant species or equivalently by compressing the target 
species, allowing heat absorption by the coolant species. Although we have focused on 
the first protocol, this should be equivalent in producing a phase space increase for the target 
species. 

The demonstration closest to shortcuts to adiabaticity in the framework of two-species 
trapping and cooling has been reported by the Florence group \cite{Catani2009}. 
A Bose-Bose mixture consisting of ${}^{41}$K and ${}^{87}$Rb is cooled to nearly quantum 
degeneracy in a magnetic trap. An optical dipole trap is then formed by using a laser 
beam operating at a wavelength intermediate between the two dominant transitions of Rb 
(the so-called D1 and D2 lines). With this choice, there is nearly complete cancellation 
of the dipole forces on Rb (the Rb potential is found to change by only 8 $\%$ with respect 
to the K potential), as discussed in \cite{LeBlanc2007}, while the trapping potential felt 
by ${}^{41}$K is stiffer, leading to its compression. The latter is carefully designed in 
such a way that it is quasi-adiabatic, because the exponentially time-dependent 
ramp has a rising time longer than the trapping periods of ${}^{41}$K atoms in the pre-existing 
magnetic trap. The experimenters compared various cooling and compression strategies and 
also explored their reversible nature via repeated cycles of compression and expansion.  
A schematic of the experiment is depicted in Fig. 11a, and the compression stages   
are shown to produce clouds of ${}^{41}$K with a progressively higher condensate ratio (Fig. 11b). 

\begin{figure*}[t]
\begin{center}
\includegraphics[width=0.43\textwidth,clip]{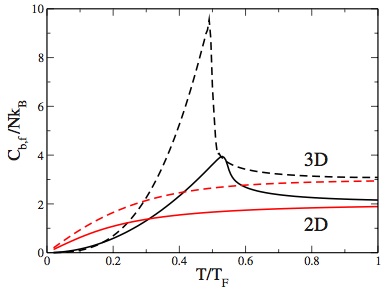}
\includegraphics[width=0.45\textwidth,clip]{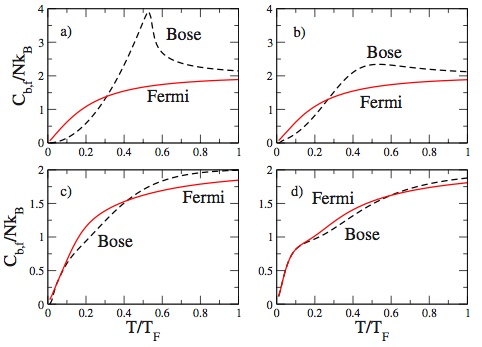}
\end{center}
\caption{Specific heat matching in lower-dimensionality systems. (Left) Heat capacity curves of bosons and fermions in 3D (dashed
curves) and 2D (solid curves) are shown versus the degeneracy parameter $T/T_F$ in the case of an isotropic harmonic trap. 
The crossover between the two curves in the 2D case occurs at a slightly higher $T/T_F$ value, ruling out its use for a more 
favorable cooling of fermions. (Right) The case of progressively one-dimensional trapping obtained by squeezing a 2D trap 
by increasing the trapping frequency ratio from $\omega_y=\omega_x=1$ (a), to $2.5 \times 10^3$ (b), $2 \times 10^4$ (c), and 
$5 \times 10^4$ (d). In the last two cases the difference between the Bose and the Fermi curves is minimal. The case of an equal number 
of bosons and fermions $N_\mathrm{b}=N_\mathrm{f}=10^4$  with equal masses $m_b=m_f$ is analyzed (reproduced from \cite{Brown2008}).}
\end{figure*}

The authors also highlight the limitations of the technique, especially with regard to 
the heating rate for ${}^{87}$Rb of the order of 0.7 $\mu$K s${}^{-1}$, reduced by one order of magnitude through 
a microwave shield. In spite of this precaution, the number of ${}^{87}$Rb atoms decreases at a rate of 
$2.5 \times 10^5$ s${}^{-1}$, and further increases during the compression stage, which is attributable to the large Rayleigh 
scattering due to the proximity of the laser wavelength to the D1 and D2 lines. 
After 5 cycles, the lower number and the higher temperature before the compression are not 
sufficient to drive ${}^{41}$K to quantum degeneracy. Shortcuts to adiabaticity techniques could allow 
a much faster adiabatic compression of the target species, reducing atomic losses in the ${}^{87}$Rb heat bath.

\subsection{Heat capacity matching by reduced dimensionality}

The ultimate physical reason for the mismatch between the specific heat of Bose and Fermi degenerate gases is related to the 
different dependences of the density of states on the energy, and this depends in turn on the effective dimensionality of the Bose gas. 
Therefore, it is important to explore the possibility of matching the heat capacity of Bose and Fermi gases at the lowest possible $T/T_F$ 
by exploiting lower dimensionality traps \cite{Brown2008}.  As discussed in \cite{Rehr1970,Bagnato1991} and experimentally demonstrated in 
\cite{Ketterle1996,Moritz1994}, a dramatic increase in the trapping frequency in one (or two) trapping axes results in an effective two- 
(or one-) dimensional system. This in turn allows a better matching of the heat capacities, because the Bose gas dependence on 
temperature becomes milder than in the full 3D case. In order to gain quantitative insights on how to realize such a matching, we 
first consider noninteracting gases at thermal equilibrium at a temperature $T$ in a harmonic potential, with the number $N_\alpha$ 
of particles fixed ($\alpha=B,F$ for Bose and Fermi gases), and expressed in terms of the chemical potential $\mu_\alpha$ as 
\begin{equation}
N_\alpha(\mu, T) = \sum_{j=0}^{\infty} \frac{g_j}{\exp{(E_j - \mu)/k_B T} \pm 1}
\end{equation}
where $g_{j}$ is the degeneracy of energy level $E_{j}$, and + (-) refers to Fermi (Bose) gases.
The expression can be  numerically evaluated and inverted to obtain the dependence of the chemical potential upon the temperature, 
$\mu = \mu (T)$, which allows evaluating the total energy 

\begin{equation}
E_\alpha(\mu, T) = \sum_{j=0}^{\infty} \frac{g_j E_j}{\exp((E_j - \mu(T))/k_B T\pm 1}
\end{equation}
and the heat capacity $C (T) = \partial E/\partial T$.  As depicted in Fig. 12, the Bose and Fermi heat capacities intersect each other at 
$T \simeq 0.293~T_F$ in the three-dimensional case, and at $T \simeq 0.308~T_F$ in the two-dimensional case, where the Fermi temperature is 
given by $T_F^{2D} = (2N_\mathrm{f})^{1/2} \hbar \omega/k_B$. Therefore, the reduction in dimensionality from three to two actually makes the heat capacity 
matching slightly worse. However, the numerical study shows that upon approaching a full one-dimensional case, the heat capacities coincide. 
This effect is better appreciated if the dimensionality is gradually reduced by progressively increasing one of the two trapping frequencies, 
$\omega_x=\omega$ and $\omega_y=k \omega$, and then gradually increasing the relative confinement parameter $k$, with the  
system becoming effectively one-dimensional when $k_B T \ll k \hbar \omega$.  As shown in Fig. 12b-e, with increasing $k$, the
shape of the heat capacity curve of bosons becomes more similar to that of fermions, in the sense that it slowly loses the peak structure  
and the curvature near the zero temperature begins to resemble that of fermions. If $k$ is increased further (Fig. 12d), there is
 a region where the two curves completely coincide with each other. This is consistent with the previous result for an ideal 1D
trap.  In the ideal 1D case, the heat capacity curves are identical, as is understandable in the canonical ensemble approach \cite{Schonhammer2000,Mullin2003}. 
Indeed, the total internal energy of fermions in a 1D harmonic trap only differs from that of the bosons by the Fermi zero-point energy $E_0 =N_\mathrm{f}(N_\mathrm{f}-1)\hbar \omega/2$, 
and therefore the two systems have identical heat capacities. 

The existence of this crossover indicates that we can control the heat capacity matching of bosons and fermions by changing 
the ratio of the two trapping frequencies in a 2D trap. Thus, one possible solution is to improve the cooling efficiency is 
first evaporating in a 3D trap and then, when the Fermi degeneracy approaching $T/T_F \simeq 0.3$, increasing the trapping 
frequencies, thus achieving a quasi-one-dimensional system and thereby continuing the evaporation process.
A possible limitation of this technique comes from the larger collisional loss rate as a result of the increased confinement.
Also, as studied for achieving Bose condensation of hydrogen atoms, the nearly 1D character of the evaporative cooling \cite{Pinkse1998} can  
lead to nonergodic evaporation limiting its efficiency \cite{Surkov1996}, although this has not prevented achieving Bose degeneracy 
\cite{Gorlitz2001a,Schreck2001b,Greiner2001,Moritz2003}. 

The discussion of lower dimensionality so far has been focused on ideal, noninteracting Bose and Fermi gases. When interatomic interactions 
are considered, various considerations can be added, especially with regard to the possibility to study conventional or unconventional superfluid phases.
As an example, the BCS pairing in Fermi-Bose mixtures in two-dimensional systems has been discussed in \cite{Mur2004}, with the maximum 
gap energy for pairing found in the case where the boson density reaches an optimal value. The observation of LOFF states also seems more favourable in two dimensions 
\cite{Samokhvalov2010} and in one dimension \cite{Dalmonte2012}. In the last paper, a theoretical comprehensive analysis was carried out for gases with spin 
inbalance, mass imbalance, and species-selective trapping, identifying Fermi-Fermi mixtures such as ${}^6$Li-${}^{40}$K as the most promising to 
observe LOFF superfluidity. So far, the only claim for a phase diagram in which a large polarized region seems to be consistent with the expectations 
from a LOFF phase has been obtained in spin-imbalanced one-dimensional fermions \cite{Liao2010}. Therefore we expect lower dimensional Fermi systems 
to play a major role anyway, and it is possible that heat capacity matching as discussed in this section will be a further reason to pursue this route. 

\begin{figure*}[t]
\begin{center}
\includegraphics[width=0.35\textwidth,clip]{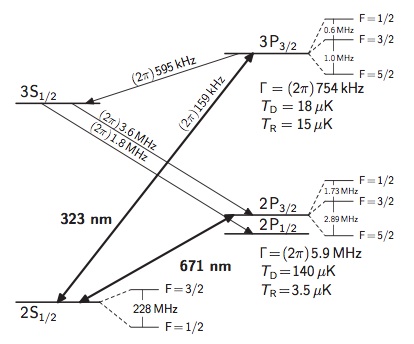}
\includegraphics[width=0.49\textwidth,clip]{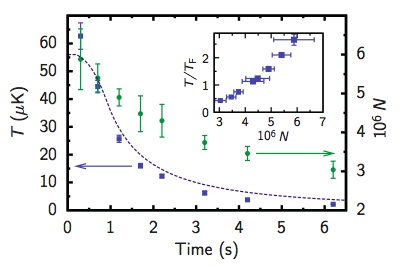}
\end{center}
\caption{Laser cooling of ${}^6$Li using the $2S_{1/2}\rightarrow 3P_{3/2}$ narrow-line transition. On the left a schematic of the 
energy levels involved and a comparison to the usual $2S_{1/2}\rightarrow 2P_{3/2}$ cooling is shown, with the indication of 
the Doppler and recoil temperatures for both transitions. On the right the dynamics of evaporative cooling are shown with the 
dependence on time of the number of atoms $N$ (green circles, right vertical scale) and the temperature $T$ (blue squares, left vertical scale). 
In the inset the trajectory in the $(N,T/T_F)$ plane is shown, with the leftmost point occurring for $3 \times 10^6$ atoms at $T/T_F=0.45$ 
 (reproduced from \cite{Duarte2011}).}
\end{figure*}

\subsection{All-optical cooling techniques}

Although evaporative cooling has enjoyed immediate and pragmatic success in the case of Bose gases in spite of its main shortcoming, the 
drastic reduction in the number of cooled atoms, many laboratories have continued to pursue number-preserving optical cooling techniques. 
Evaporative cooling is less universal than optical cooling techniques in general, as it relies on the dominance of  elastic over inelastic 
collisions and all sources of energy-unselective loss of atoms, such as three-body recombination and dipolar relaxation, 
and this depends on the specific details for each atomic state in each atomic species. With the advent of ultracold Fermi gases and the associated 
Pauli blocking effect, the limitations of evaporative cooling became even more evident. This has led to the development of all-optical cooling 
techniques specialized for fermions. From the instrumental standpoint, a single-atom cooling technique (usually as optical cooling can be considered) 
also benefits from the simplification in not needing a second species as in sympathetic cooling. Additionally, the duty cycle of a purely 
optical cooling scheme is usually faster than when evaporative and sympathetic cooling stages are involved, because there is no need for thermalization among atoms. 
We can classify recent advances in this subfield into two categories. In the first, the idea is to act on 
improving the precooling stage at the level of magneto-optical trapping and cooling.  This creates more favourable conditions for 
an efficient transfer of cold atoms from the MOT to the optical dipole trap, without the need of magnetic trapping as an intermediate stage. 
In the second, optical cooling is considered as an efficient tool all the way down to quantum degeneracy. 

The need for an intermediate stage of magnetic trapping, resulting in complications in the apparatus, including optical access to the atomic cloud 
for manipulation and imaging, emerges due to the limited trap depth of even high-power optical dipole traps. This generates a temperature gap between 
the minimum temperature achievable in a MOT and the maximum one affordable for efficient transfer, with minimal losses, in the optical dipole trap. 
With the advent of higher-power laser sources, especially due to progress in solid-state lasers, this gap is being reduced. However,  
it is not possible to use sub-Doppler cooling for the only stable alkali-metal atoms of fermionic nature, ${}^6$Li and ${}^{40}$K. 
Three groups have overcome this limitation by exploiting a cycling transition for the MOT corresponding to a linewidth of the upper state smaller than 
for conventional MOTs, also named UV-MOT due to the use of smaller laser wavelengths. More specifically, the Rice group has used the $2S_{1/2}\rightarrow 3P_{3/2}$ 
transition in ${}^6$Li  following the usual MOT cooling based on the $2S_{1/2}\rightarrow 2P_{3/2}$ transition \cite{Duarte2011}. 
This results in temperatures at the end of the UV-MOT stage as low as 59 $\mu$K, about three times lower than the Doppler limit based 
on the $2S_{1/2}\rightarrow 2P_{3/2}$ transition. After 6 seconds of dual evaporative cooling, $3 \times 10^6$  atoms were brought to the 
Fermi degeneracy $T/T_F=0.45$ (Fig. 13). A group at the University of Toronto has used the analogous transition in ${}^{40}$K, $4S{1/2} \rightarrow 5P_{3/2}$, 
following conventional MOT cooling based on the $4S{1/2} \rightarrow 4P_{3/2}$ cyclic transition \cite{McKay2011}, experiencing a gain in reduced temperature of 
about a factor of four compared to the $4S{1/2} \rightarrow 4P_{3/2}$ transition.  Narrow-line cooling of ${}^6$Li has also been studied in detail with regard to the 
factors limiting the density and the associated light-assisted losses, achieving a record temperature in the UV-MOT stage of 33 $\mu$K, less than a factor two from 
the theoretical expectation for the Doppler temperature of the $2S_{1/2}\rightarrow 3P_{3/2}$ transition \cite{Sebastian2014}. 
Groups at the ENS \cite{Fernandes2012} and at LENS in Florence \cite{Burchianti2014} have instead used so-called gray-molasses cooling with 
significant simplification in terms of the MOT setup, avoiding the need to use near-UV lasers and broadband optical components. 
In the latter case, the precooling stage also involved a velocity-selective coherent population trapping stage. After dual evaporative in an optical dipole 
trap, $3.5 \times 10^5$ atoms were produced at a Fermi degeneracy factor $T/T_F \simeq 0.06$. 

\begin{figure*}[t]
\begin{center}
\includegraphics[width=0.48\textwidth,clip]{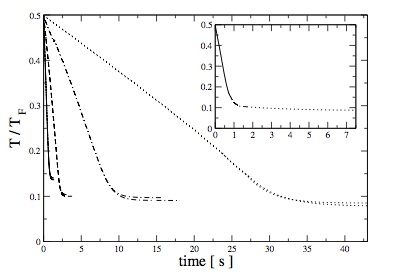}
\includegraphics[width=0.49\textwidth,clip]{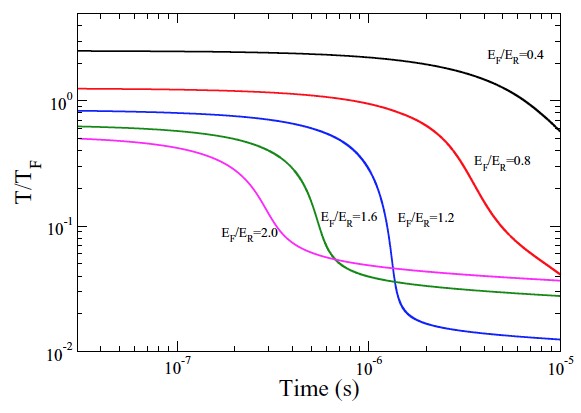}
\end{center}
\caption{Examples of proposals to achieve deep Fermi degeneracy with atom number-preserving all-optical methods. 
On the left, the Fermi degeneracy factor of a gas of ${}^{40}$K versus time is shown for various values of the spontaneous emission 
rate $\gamma$ and the effective interatomic coupling constant $g$, as defined in \cite{Dziarmaga2005a}, with laser cooling 
via a series of Raman pulses on the attractive side of a Feshbach resonance (reproduced from \cite{Dziarmaga2005a}). 
On the right, the same quantity is evaluated for different values of the initial $T/T_F$ and $E_F/E_R$ in the case of runaway cooling 
in a combined MOT-ODT trap resulting from the progressively reduced Doppler temperature. The cooling rate is low both initially, when the 
atoms have low quantum degeneracy, and at later times, when the Pauli principle inhibits the dynamics (reproduced from \cite{Onofrio2016}).}
\end{figure*}

The abovementioned cooling techniques are instrumental in avoiding the use of a magnetic trap, but are not meant to bring the atoms to degeneracy 
with optical cooling, the last stage always consisting in forced evaporative cooling from the optical dipole trap. More radical solutions have been envisaged in 
which fermions are cooled to degeneracy by purely optical means. In this framework, a distinctive feature of Fermi gases is that inhibition of the spontaneous 
emission is expected in the degenerate regime \cite{Helmerson1990,Imamoglu1994,Javanainen1995,Busch1998,Gorlitz2001b,Shuve2010}, 
and this is considered a drawback because it would slow the cooling dynamics in a MOT. 

In a series of papers, Lewenstein and collaborators \cite{Idziaszek2001,Idziasek2002,Idziasek2003,Dziarmaga2005a,Dziarmaga2005b} proposed a cooling 
scheme in which atoms are confined in an optical dipole trap, and trains of pulses for Raman transitions  are generated. This creates an effective two-level 
system after integrating out the dynamics of the intermediate upper level in the Raman transition. The effective Rabi frequency and the effective 
spontaneous lifetime are controlled by the experimenter by adjusting the power and timing of the Raman pulses. Photon reabsorption and inhibition of the 
spontaneous emission are therefore under control and various simulations, including realistic heating sources, show that Fermi degeneracy factors $T/T_F < 0.1$ 
can be achieved. In Fig. 14a, the cooling dynamics are depicted for various values of the effective spontaneous linewidth $\gamma$, showing that 
Fermi degeneracy factors $T/T_F \simeq 0.08$ are achievable.  

A different route to achieve Fermi degeneracy with MOT cooling has been proposed in \cite{Onofrio2016}.  
In this case, the very inhibition of spontaneous emission is exploited to obtain lower $T/T_F$ based on the 'runaway' effect in a combination of a MOT  
and an optical dipole trap. If the atomic cloud already starts with a small but finite degree of quantum degeneracy, the Doppler cooling limit in a magneto-optical trap 
tends to a smaller value, which in turn enhances the degree of quantum degeneracy. A positive feedback mechanism can then be ignited in such a way 
that the Doppler limit reaches its limiting value at the recoil temperature. 
The required initial Fermi degeneracy can be achieved by increasing the Fermi temperature using an optical dipole trap. 
The combination of a MOT for cooling and an optical dipole trap with a variable trapping frequency should ensure fast 
cooling, and estimates indicate that the cooling and heating rates, at least if the dominant source for the latter is Rayleigh scattering, should 
should equalize at about $T/T_F \simeq 10^{-2}-10^{-1}$. Given the stiff confinement in the optical dipole traps used in \cite{Duarte2011,Burchianti2014}, 
it is possible that a regime of inhibition of spontaneous emission was present before starting evaporative cooling. Interestingly, the Rice group 
reported improved loading when the UV-MOT was kept on for 5 ms following the start of the ODT stage, and the Florence group 
commented on the efficiency of Doppler cooling after transfer in a deep optical dipole trap. In a recent paper \cite{Gross2016}, the Singapore group  
systematically studied the persistence of Doppler cooling in the ODT. The number of atoms in the ODT was found to be directly proportional to the duration of the 
stage in which both the UV MOT and the ODT are present, up to 10 ms, and thereafter slowly decreases, presumably due to density-dependent losses. An estimate of 
the parameter space shows that in this experiment the atoms in the combined MOT-ODT have a Fermi degeneracy parameter $T/T_F \simeq 2.8$ and a 
Fermi-energy to recoil-energy ratio $E_F/E_R \simeq 1$, rather favorable for initiating the Pauli cooling mechanism discussed in \cite{Onofrio2016}.

\section{Optical lattices}

Trapping and cooling of atoms in single-minimum potentials is often considered as a preliminary stage before loading the atoms into optical lattices, in which 
multiple minima of the potential are generated in a regular pattern. This mimics crystal lattices in Nature, but with controllable 
parameters, including the geometry of the crystal itself, the distance between the minima, the potential depth, the interatomic interaction, the 
presence of defects or impurities, and time-dependent modulation of the potential. 
These systems have the potential to act as analog computers (also currently called 'simulators') in which the task is to isolate some specific effects of 
more complex condensed matter systems, allowing a purer comparison between theory and experiment in terms of solving specific model Hamiltonians.
Due to the tight confinement achievable in optical lattices and the consequent stronger interatomic interactions, optical lattices have also been  
considered to be a shortcut compared with single-trap configurations for the observation of fermion BCS-like superfluidity \cite{Hofstetter2002}.

Mixtures of fermions and bosons are intriguing since demixing, Mott-insulator, and disordered phases have been predicted \cite{Albus2003}.
Potentially, nearly all phenomena already investigated in crystal physics could be studied with these analog computers, but for quintessential quantum 
phenomena, the practical feasibility is often limited by temperature effects. This means that, again, the degree of quantum degeneracy actually achievable 
in the laboratory is often the bottleneck for the full exploration of this subfield. Several review papers and books have already  appeared discussing possible 
physics explored and explorable in optical lattices (see, e.g., \cite{Bloch2008,Lewenstein2012}). More specific to our discussion, a dedicated 
review has been devoted to the issue of cooling techniques in optical lattices, including a discussion of the related thermometry \cite{McKayRev2011}.

Rather than simply quoting this excellent review, apart from a succinct summary, we here discuss progress since 
its appearance, especially in relation to the cooling techniques already discussed in the preceding section. The cooling techniques discussed in 
\cite{McKayRev2011} can be separated into two categories, basically mimicking the single-trap case already discussed. In analogy to dual 
evaporative cooling, selective removal of the hottest atoms, which McKay and DeMarco call 'filter cooling', can be achieved in an optical lattice. 
One option is to properly shape the confinement, creating a gapped phase in the center of the trap, where the atoms have lower entropy, thereby separating 
this from atoms in the higher-entropy region at the edge, which are then removed. Filtering can also occur in energy space by removing atoms belonging to higher bands 
or transferring atoms with higher number fluctuations into another internal state. These techniques, respectively named spatial, band, and number filtering, 
require protocols in which the trapping potential is changed in time, potentially creating heat. 

The second class of cooling techniques, named immersion cooling, is analogous to sympathetic cooling, and consists in exposing the target Fermi system 
to a coolant system, for instance, a large-heat-capacity Bose gas not necessarily trapped in the optical lattice, just simply harmonically trapped. 
Here again, a Bose refrigerator experiences the same limitations in heat capacity matching as outlined in the preceding sections. 
Furthermore, its presence can complicate the analysis of the dynamics of the Fermi atoms in the optical lattice, apart from indirectly inducing heating, for instance, 
in the case of optical dipole trapping. A systematic analysis of heating and cooling in optical lattices for atomic mixtures is offered in \cite{LeBlanc2007}, with the 
identification of optimal configurations for the laser wavelengths creating species-specific optical lattices. The role of heating sources in determining the minimum 
achievable temperature comparing the heating and cooling rates is discussed in \cite{McKayRev2011}.  

Unfortunately, even before considering heating sources related to specific cooling protocols, the atoms must first be transferred from a single trap into the optical 
lattice, and ramping up the optical lattice potential would in general induce heating. Its effect is more pronounced for loading procedures with 
extended duration due to the continuous presence of heating sources, such as Rayleigh scattering due to the laser beams shaping the optical lattice potential. 
Therefore it is advisable to achieve loading in the shorter timescale still compatible with the minimal perturbation of the temperature and entropy of the atoms. 
This is a delicate interplay and nonholonomic coherent control techniques \cite{Lloyd1995,Harel1999,Brion2006} have been adapted to minimize heating 
via nonadiabatic loading \cite{Liu2011}.  The authors show experimentally, by measuring the contrast of the interference fringes after releasing the atoms 
from the lattice, that a properly designed sequence of pulses delivered to the atoms while still in the harmonic trap and prior to their loading into the optical 
lattice reduces the level of excitation and heating. The pulse sequence is chosen to maximize the fidelity in the targeted ground state of the optical lattice. 
Various patterns differing in the number of free parameters, for instance, freezing the amplitude of the pulses to be constant, have been tested, resulting 
in loading on a timescale of a few dozen microseconds. Although the analysis is made explicit for a BEC of ${}^{87}$Rb, due to the 
single-particle nature of the proposal, analogous considerations should also hold for degenerate Fermi gases. 

A second method, fast frictionless cooling, has also been proposed as a possible way to load atoms into an optical lattice with minimal heating. 
In a preliminary study, a fast frictionless expansion of atoms already in an optical lattice with a dynamically variable spacing was discussed \cite{Yuce2012}. 
The emphasis here is on combining the advantages of an optical lattice with small spacing, such as the large hopping rates available due to the proximity of 
the lattice minima, with the ones of large-spacing optical lattices, such as the easier imaging and addressability of each lattice site with usual optical techniques. 

This discussion was followed by a proposal \cite{Masuda2014} in which nonadiabatic loading of a BEC with high fidelity is also achievable 
on the timescale of a few dozens of microseconds using the fast-forward approach \cite{Masuda2008,Masuda2010,Masuda2011}. 
The technique is reminiscent of but not equivalent to fast frictionless cooling, because antitrapping stages are also present in this case. 
The authors discuss both implementations using painting potentials achieved through fast time-averaged optical potentials 
\cite{Onofrio2000b,Milner2001,Friedman2001,Henderson2009} or bichromatic optical lattice potentials combining 
red-detuned and blue-detuned laser beams, the optical lattice counterpart of the bichromatic traps discussed earlier in section VI A. 
At the shortest timescale of 6.9 $\mu$s discussed in \cite{Masuda2014}, the fidelity in loading the BEC into the ground state of the 
optical lattice is expected to be above 0.95 for a fast-forward modulation of the optical lattice potential, systematically above the one achieved by using time modulation based on fast frictionless cooling, as discussed in \cite{Yuce2012}.

In a third paper \cite{Dolfi2015}, loading a one-dimensional Bose gas into an optical lattice was simulated numerically in the continuum. The leading source of heating was 
found to be the atomic density redistribution, dominating by at least an order-of-magnitude competing sources of heating, such as the nonadiabatic heating into higher bands. 
Both the homogeneous case (no harmonic trapping) and the more realistic inhomogeneos case have been discussed. The inhomogeneous case corresponding to harmonic trapping leads to 
significantly stronger heating than in the homogeneous case, and the authors propose simple time modulations of the trapping frequency, even just a linear chirp, which
are very effective in reducing heating by more than an order of magnitude. 

\begin{figure*}[t]
\begin{center}
\includegraphics[width=0.45\textwidth]{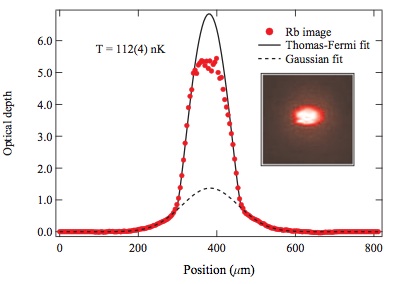}
\includegraphics[width=0.45\textwidth]{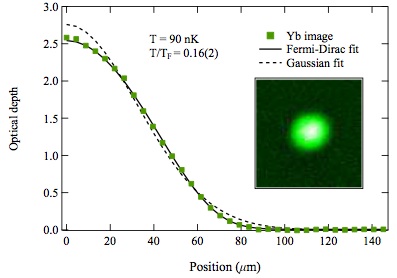}
\caption{Precision thermometry in a quantum degenerate Bose-Fermi mixture. The images show atomic cloud profiles from absorption imaging 
of ${}^{87}$Rb (left, $5 \times 10^5$ atoms, time-of-flight of 22 ms) and ${}^{171}$Yb (right, $2.4 \times 10^5$ atoms, time of flight 12 ms). 
The temperature of the ${}^{87}$Rb cloud is determined by fitting the wings of the profile containing the thermal component alone with a 
Gaussian function,  yielding a temperature of 112 nK, with a relative error estimated to be 3.5 $\%$. The fit of the ${}^{171}$Yb with a Fermi-Dirac 
distribution is more accurate than a (thermal cloud) Gaussian fit and yields a temperature of 90 nK with a relative error of 12.5 $\%$ 
(reproduced from \cite{Vaidya2015}).}
\end{center}
\end{figure*}

Finally, we mention the possibility to have built-in sources of cooling during the optical lattice stage. A first proposal in this direction was discussed 
in \cite{Mathy2012}. Three laser beams with different waists and the opposite detuning with respect to the beams producing the optical lattice are superimposed 
on the optical lattice. The resulting antitrapping contribution keeps the chemical potential of the gas in the gap of the interested phase, for instance,  
a Mott insulating phase. Stages of evaporative cooling in optical lattices could be a solution to the increase in $T/T_F$ otherwise expected after loading 
from a single trap and an alternative to the use of immersion cooling techniques based on the continuous presence of a Bose refrigerator. More recently, addition of 
a disordered potential to an optical lattice and the tracking of a constant-entropy trajectory have been suggested as a cooling mechanism \cite{Paiwa2015}.
However, as discussed in the concluding remarks of \cite{Paiwa2015}, this setup requires a concurrent cooling system to decrease the entropy in 
the initial disordered lattice, for instance, as proposed in \cite{Mathy2012}. As an alternative, it may be possible to exploit  runaway cooling,  
as discussed at the end of Section VI D, because optical lattices allow confinement strengths corresponding to rather large Fermi temperatures. 
Three promising steps in this direction have been the observation of suppression of recoil heating of a Bose gas \cite{Wolf2000}, a theoretical study 
of the inhibition of spontaneous emission \cite{Sandner2011}, and the recent observation of microscopic effects of Pauli blocking \cite{Omran2015}, all 
related to optical lattice configurations.

\section{General problems in precision thermometry}

The study of phase transitions in ultracold fermions, unless quantum phase transitions are considered - albeit often occurring at very low or inaccessible 
temperatures - requires mapping the phase diagram, and it is therefore crucial to assess the temperature of the atomic cloud. In the case of a Bose gas, this is 
readily available because at any finite temperature there are atoms in the thermal component superimposed on the Bose condensed component.  
In a strongly interacting Bose gas, the Bose-condensed component has a Thomas-Fermi spatial profile that is quite different from the nearly 
Gaussian profile of the thermal component. The two components can then be easily discriminated for instance in time-of-flight absorption imaging. 

By contrast, the temperature of a Fermi gas alone is difficult to assess due to the weak dependence of many properties at very low temperature. 
This is intrinsic to the fermionic nature of the gas, because at very low temperature the average occupation number per state is almost approximated as a 
Heaviside function with very little dependence on temperature, a property responsile, for instance, for the small contribution to the specific heat or to 
the Pauli paramagnetism in metals. The situation is summarized with the example shown in Fig. 15, where the spatial profiles of a Bose and a Fermi gas 
arising from the same mixture are shown. While the Fermi-Dirac distribution yields a better fitting of the spatial profile compared with the classically expected 
profile, their difference is not striking if compared with the spatial profiles of the two components of the Bose gas. It is then natural, at least in the case of 
Bose-Fermi mixtures, to assess the temperature based on the fitting of the wings of the Bose distribution. Extrapolating this measurement to the Fermi gas 
assumes thermal equilibrium between the two species, which may be invalid, especially if the interspecies elastic scattering length is small, as for instance in the  
case of ${}^{6}$Li-${}^{87}$Rb mixtures \cite{Silber2005}. Also, this shows a further advantage of the bichromatic optical dipole trap dscussed in section VI A, 
{\it i.e.}, the fact that keeping the Bose gas less degenerate or even above the critical temperature for Bose-Einstein condensation as in the example 
of fast frictionless cooling in section VI B, results in a more accurate fit of the Gaussian profile and therefore of the gas temperature. 
The demonstration of the inhibition of spontaneous emission scattering for a Fermi gas 
\cite{Helmerson1990,Imamoglu1994,Javanainen1995,Busch1998,Gorlitz2001b,Shuve2010} provides an alternative way of directly measuring the Fermi degeneracy 
parameter $T/T_F$. 

With regard to optical lattices, several thermometry techniques have been proposed, and some of them have also been  demonstrated experimentally. 
A classification of possible thermometry methods prior to 2011 are discussed in \cite{McKayRev2011}. While we refer the reader to this monograph for earlier 
developments in this area, we explicitly mention explicitly the pioneering discussion reported in \cite{Kohl2006}. The author, after having quantified the increase 
in $T/T_F$ when adiabatically loading the atoms from a harmonic trap into an optical lattice, proposed the fraction of doubly occupied lattices sites as a sensitive temperature 
indicator in the case of a two-component Fermi gas. This indicator can be measured by determining the number of molecules produced in the strong coupling limit 
provided by a Feshbach resonance. Other thermometry methods recently proposed involve the determination of the width of the intermediate magnetization region for two 
spatially separated spin-polarized components \cite{Weld2009,Weld2010},  the measurement of spatially resolved density and the related number fluctuations \cite{Zhou2011}, 
Raman spectroscopy \cite{Bernier2010}, light diffraction \cite{Ruostekoski2009} and light shifts \cite{McDonald2015}, the response of the Fermi gas to an 
artificial gauge field \cite{Roscilde2014}, and lattice amplitude modulation \cite{Loida2015}.
The general issue with many of these measurement techniques, as emphasized in \cite{McKayRev2011}, is that a network of benchmarks is required for calibration purposes, 
and preferably they should not be model-dependent, or at least they should not rely on the same physics to be explored to avoid circularity.

Under these circumstances, it is preferable that future experiments determine the temperature in at least two different ways, to verify the internal 
consistency and properly assess temperature error bars. Furthermore, it would be important to design a noninvasive measurement technique to allow repeated monitoring 
of the same system along a dynamical phase transition, also still an open problem considering the level of perturbation introduced in measuring such cold 
and tiny atomic samples.

\section{Conclusions}

We have discussed cooling techniques and methods to perform precision thermometry for ultracold Fermi gases. 
The review has been developed in three related areas: providing some physical motivations, describing the experimental 
setups capable of achieving these physics goals, and discussing a variety of techniques to overcome or mitigate limitations in achieving these goals. 

It is also important to remark at this point that this review does not exhaust the many proposals or ongong research directions, and 
has to be considered a critical stimulus to identify some fields that appear more promising for fully enabling the potential of 
this fascinating subfield of atomic physics. After all, the highest aspiration of a review paper is to quickly become obsolete due to the work 
generated by its critical reading. Indeed, some of the considerations reported in sections VI-VIII lend themselves to opening up 
novel experimental programs, as some demonstrations have been implemented only on Bose gases.  
In general, ultracold atoms seem ideally suited to study nonequilibrium statistical physics, considering the control over  
timescales and lengthscales, geometries, and interaction strengths. It is possible that, regardless of the many promises which cannot be 
fulfilled in the future, this research avenue is going to stay with us for a long time, influencing our conceptual understanding of how 
statistical systems reach, or fail to reach, equilibrium states. Finally, we want also to take this opportunity to address more general 
questions any curious reader could raise, like the following.

\vspace{2.0mm}

\noindent
$\bullet$ {\sl Do we need optical lattices for most of the ultracold atomic physics we want to explore?}

We have seen that achieving deep Fermi quantum degeneracy seems to be difficult, also in light of the many concatenated protocols and 
transfer from a single trap to the multitrap configuration. A sensitive question may be if we need to necessarily 
struggle with optical lattices in the first place. One of the motivations usually put forward for their need is that they provide a possible platform 
for quantum computers, but we find this motivation rather weak.  The scalability of optical lattice setups, at least compared with hardware based on 
solid-state devices, is rather questionable and, at the demonstration level, ion traps are already outperforming optical lattices. 
To pursue optical lattices merely based on quantum information motivations seems prone to criticism and also potentially dangerous, because it 
could result in a dead end, considering the ongoing fierce competition among several experimental techniques. Optical lattices also have various 
issues compared with real crystals, such as the presence of quasiperiodicity due to the coexistence of the optical lattice potential with 
a (usually weaker) harmonic confinement, which breaks the translational invariance and complicates the analysis in terms of periodic 
wave functions. Due to this specific feature, we believe that a possible feasible contribution of optical lattices in the future could be 
in addressing surface physics problems, which are typically difficult to handle in solid-state physics. Morever, there are features of 
phase transitions that could already be explored in single-minimum or few-minima traps, avoiding the complications of optical lattices and 
the necessarily higher Fermi degeneracy factors without dedicated cooling stages. As an example, the Mott-superfluid phase transition 
successfully investigated in optical lattices for Bose gases \cite{Greiner2002} and Fermi gases \cite{Jordens2008,Schneider2008} 
may have a physical counterpart in the observation of self-trapping and tunneling in a double-well trap \cite{Albiez2005}. 
In any event optical lattices are used as emulators of the actual condensed matter physics problem, and it may be therefore advantageous 
to work with simpler emulators in which lower temperatures are achievable, at the price of complicating the theoretical analysis 
bridging the gap between the emulator and the actual system. 

The analysis of phase transitions, usually considered to be well-defined in the thermodynamic limit, requires a heavy use of scaling 
arguments to disentangle finite-size effects in single-minimum or few-minima trap configurations.  In addition, energy bands central to the physics 
of long-range-order solids are difficult to mimic in a single-minimum or few-minima traps. An interesting compromise between optical lattices 
and single- or few-minima traps may be the systematic use of ring geometries, so far limited to Bose gases \cite{Ramanathan2011,Moulder2012,Ryu2013,Edward2015}, but 
the performance of this approach in terms of cooling capability has yet to be explored. 

\vspace{2.0mm}

\noindent
$\bullet$ {\sl Are ultracold atoms going to replace the study of condensed-matter systems?}

As already outlined in this review, ultracold atoms present several advantages in studying many-body systems, especially in tracking the detailed dynamics 
due to the slower timescales and the larger lengthscales on which nontrivial phenomena, such as the formation and decay of vortices, occur. 
Furthermore, the systems are rather pristine since impurities or 'minority' atoms - at least for the typical operating vacuum of the experiments - are either absent, 
due to the selective spectroscopic preparation in the precooling stage, or controllable, for instance using electromagnetic pulses to convert atoms in another 
hyperfine state. However, we always have to keep in mind that they are artificial, human-made systems that are always in a metastable state. 
Phenomena such as persistent currents can still be explored, but not at the same level of accuracy and duration as already done in superfluids and superconductors. 
In spite of a steady progress in laser technology, ultracold atoms laboratories are expensive and require substantial building effort, also combining 
several technologies pushed together to the ultimate limit, such as ultra-high vacuum and high-resolution manipulation and detection of atoms clouds. 
The advantage of available flexibility in setting up various trapping configurations using magnetic fields and lasers is counterbalanced by the fact 
that in condensed matter physics there are a huge number of samples, both available in Nature and human-made, which can be used to understand 
basic phenomena, often just simply changing the sample in the apparatus. Also, the typical Fermi quantum degeneracy factors $T/T_F$ 
in condensed-matter systems are much smaller than in ultracold atoms, due to the large Fermi energy of lighter particles such as electrons 
naturally confined at densities typical of solid or liquid state materials. In mK or $\mu$K condensed-matter systems, continuous cooling techniques are 
well established, with minimal or zero heating rates during the preparation and measurement stages, and most importantly without particle loss or 
changes in the sample parameters. 

In spite of these disadvantages, the study of the unitarity regime for Fermi gases and the BEC-BCS crossover is a major interdisciplinary contribution of 
ultracold atomic physics with no equivalent in condensed-matter physics, or studies of quark matter in nuclear, particle physics and astrophysics. 
In light of all these considerations, a balanced research program on Fermi systems should perhaps consider a plurality of experimental approaches.
This implies including more established techniques in the research agenda to study actual superconductors and superfluids at the condensed-matter level, 
also exploiting the easier access to observable quantities due to the charged nature of electrons as typical probes of transport phenomena. 
The promise of capturing the essential features of high-temperature superconductors via model Hamiltonians that can be simulated via analog computers 
made of ultracold atoms still has to be fulfilled, and various hurdles on the road could limit its potential, and hence in the absence of further breakthroughs in 
cooling techniques it seems reasonable to take a more cautionary approach. 

\vspace{2.0mm}

\noindent
$\bullet$ {\sl Are there fundamental limits to the minimum achievable Fermi degeneracy?}

All considerations reported in this review assume that the temperature of the cloud, even if very low, is still higher than the one corresponding to
the energy level spacing of the harmonic oscillator schematizing the trapping potential. Eventually, the temperature becomes close to the ground 
energy of the harmonic oscillator, and this establishes a fundamental limit, unless specific techniques are adopted to squeeze thermal fluctuations.
Therefore we expect the 'standard quantum limit' for a gas to yield an upper bound to the achievable temperature in the range of 
$T_{\mathrm{min}} \simeq \hbar \omega/(2 k_B)$, for an average trapping angular frequency $\omega$, on which the Fermi temperature 
also depends, $T_F = \hbar \omega (6N_\mathrm{f})^{1/3}/k_B$, and we therefore obtain a limit on the minimum achievable 
Fermi degeneracy factor 

\begin{equation}
\frac{T_{\mathrm{min}}}{T_F} \simeq 0.28 N_\mathrm{f}^{-1/3},
\end{equation}
regardless of the trapping potential strength as long as it is schematizable as harmonic. The ultimate limit is then established by the number of 
fermions alone, and for $N_\mathrm{f} = 10^6$ already achieved in various experiments, this implies $T_{\mathrm{min}}/T_F \simeq 2.78 \times 10^{-3}$, just one 
order of magnitude smaller than the values currently achieved!
This limit may be pushed to lower values by increasing the number of fermions, which requires collecting a large number of fermions already 
at the MOT stage, by changing the dispersion relationship for the particle via a proper choice of the trapping potential, or by using lower-dimensionality 
systems.

For instance, in two dimensions, where the Fermi energy for a harmonic trap is $E_F= \hbar \omega (2N_\mathrm{f})^{1/2}$, the corresponding minimum 
achievable Fermi degeneracy factor is $T_{\mathrm{min}}/T_F=(2 \sqrt{2 N_\mathrm{f}})^{-1}$, which implies, for $N_\mathrm{f} = 10^6$ as in the three-dimensional 
example above, $T_{\mathrm{min}}/T_F \simeq 3.5 \times 10^{-4}$, {\it i.e.} a gain of almost one order of magnitude. This gain is even more 
pronounced in one dimension, since in this case $E_F= \hbar \omega N_\mathrm{f}$, leading to $T_{\mathrm{min}}/T_F \simeq 5 \times 10^{-7}$. 

Two-dimensional Fermi gases are currently under intense experimental investigation (see \cite{Barmashova2016} for a recent review).
An intriguing possibility of circumventing the limit shown in Eqn. (7) is available by squeezing quantum fluctuations at one phase of the 
harmonic oscillator representing the motion of each atom by using time-modulated trapping. It is worth mentioning that momentum-squeezed 
states for fermions have already been demonstrated \cite{Wang2011}. Pursuing this research avenue could lead to a novel path to study quantum 
measurements in atomic many-body systems, with potential implications for measurements of fundamental physics, and in general advancing quantum 
metrology of translational degrees of freedom \cite{Braginsky1967,Braginsky1974,Braginsky1996}.

\begin{acknowledgments}
It is a pleasure to thank various colleagues with whom, during more than a decade, I have collaborated pursuing some of the research directions 
described in this review, in chronological order of collaboration: Carlo Presilla, Robin Cot\`e, Eddy Timmermans, Michael Brown-Hayes, Woo-Joong Kim, 
Qun Wei, Stephen Choi, and Bala Sundaram. I am also grateful to Emilio Cobanera, Michael K\"ohl, Giuseppe C. La Rocca, Bala Sundaram, and Kevin C. Wright for a critical 
reading of the manuscript.  I also wish to dedicate this review to the memory of the founding father of quantum metrology, Vladimir Borisovich Braginsky.

\end{acknowledgments}

\end{document}